%% file: guardrail-selection.tex
\documentclass{egpubl}
\usepackage{eurovis2026}

\SpecialIssuePaper    
\CGFccby
\usepackage[T1]{fontenc}
\usepackage{dfadobe}  
\usepackage{enumitem}
\usepackage{fancyhdr}
\usepackage{cite}  
\BibtexOrBiblatex
\electronicVersion
\PrintedOrElectronic
\ifpdf \usepackage[pdftex]{graphicx} \pdfcompresslevel=9
\else \usepackage[dvips]{graphicx} \fi

\usepackage{xcolor}
\definecolor{quoteColorCS}{HTML}{e6ebf0}
\newcommand{\hlc}[1]{{\sethlcolor{quoteColorCS!50!white}\hl{#1}}}
\newcommand{\hquote}[1]{\hlc{``\textit{#1}''}}

\usepackage{egweblnk}

\usepackage{review}
\usepackage{multicol}
\usepackage{booktabs}
\usepackage{tabularx,multirow}
\usepackage{subcaption}
\usepackage{xspace}

\usepackage{caption}
\usepackage{xcolor}

\definecolor{colNone}{HTML}{9E9E9E}                
\definecolor{colSuperData}{HTML}{57C66F}           
\definecolor{colPercentiles}{HTML}{3AA0FF}         
\definecolor{colPercentileClosest}{HTML}{E67E22}   
\definecolor{colCluster}{HTML}{e6cb38}             
\definecolor{colMetadata}{HTML}{D94BBD}            
\definecolor{colReviewGreen}{HTML}{7FB800}          
\definecolor{rowgray}{RGB}{245,245,245}

\colorlet{colPercentilesDark}{colPercentiles!80!black}
\colorlet{colPercentileClosestDark}{colPercentileClosest!80!black}
\colorlet{colClusterDark}{colCluster!80!black}
\colorlet{colMetadataDark}{colMetadata!80!black}
\colorlet{colRandomDark}{colSuperData!80!black}
\colorlet{colControlDark}{colNone!70!black}

\newcommand{\Percentiles}{\sethlcolor{colPercentiles!15!white}\hl{Percentile Markers}\xspace}
\newcommand{\PercentileExemplars}{\sethlcolor{colPercentileClosest!15!white}\hl{Percentile-based Exemplars}\xspace}
\newcommand{\Clusters}{\sethlcolor{colCluster!15!white}\hl{Cluster Representatives}\xspace}
\newcommand{\SemanticSim}{\sethlcolor{colMetadata!15!white}\hl{Exemplars with Semantic Similarity}\xspace}
\newcommand{\RandomExemplars}{\sethlcolor{colSuperData!15!white}\hl{Random Exemplars}\xspace}
\newcommand{\Control}{\sethlcolor{colNone!15!white}\hl{No Guardrail}\xspace}

\title[Guardrail Selection in Line Charts to Contextualize Persuasive Visualizations]%
      {Guardrail Selection in Line Charts \\ to Contextualize Persuasive Visualizations}

\author[K. A. Nadib, M. Kogan, A. Lex, \& M. Lisnic]
{\parbox{\textwidth}{\centering 
K. A. Nadib$^{1}$\orcid{0009-0006-7940-501X}
M. Kogan$^{1}$\orcid{0000-0002-9200-6914} 
A. Lex$^{2,1}$\orcid{0000-0001-6930-5468}
M. Lisnic$^{3}$\orcid{0000-0001-5329-4274}}\\
{\parbox{\textwidth}{\centering 
$^1$University of Utah, USA\\
$^2$Graz University of Technology, Austria\\
$^3$Worcester Polytechnic Institute, USA}}
}

\begin{document}

\teaser{
\vspace{-1cm}
    \includegraphics[width=\linewidth]{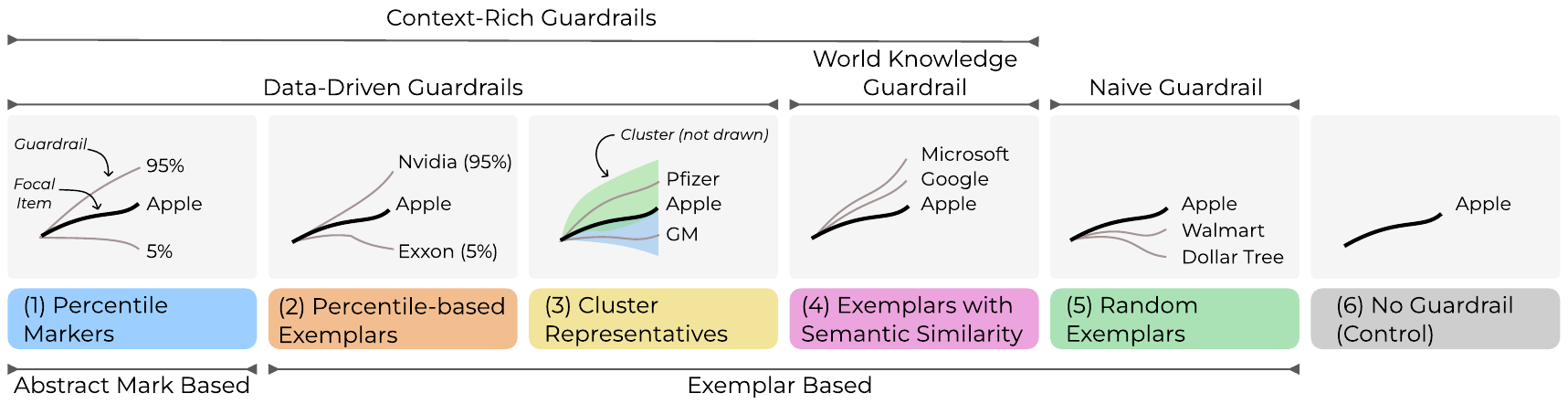}
    \caption{The five guardrails and the control condition evaluated in our experiment. We test different versions of statistical guardrails: (1) percentile markers, i.e., explicitly showing percentiles of the whole dataset over time; (2) percentile-based exemplars, i.e., showing concrete items that are close to a percentile value, (3) cluster representatives, i.e., showing items central to the clusters of the dataset, (4) exemplars with semantic similarities, i.e., based on higher-level knowledge about the data items; (5) randomly drawn exemplars, and (6) no guardrails.}
    \label{fig:teaser}
}

\maketitle
\begin{abstract}
   Charts used for persuasion can easily veer into being outright misleading when, for instance, cherry-picked data is paired with a deceptive caption, as is commonly encountered on social media. The rise of interactive time-series data explorers for hotly debated topics makes such framing easy to produce and spread. Post-hoc interventions like fact-checking often arrive too late and suffer from persistence of belief. Prior work suggests that guardrails, in the form of contextual comparison lines embedded directly into charts, can reduce these effects. We propose and evaluate a practical set of guardrail sampling strategies for implementing such contextual lines in real systems. In a preregistered mixed-design study with two real-world scenarios (COVID-19 and Stocks), participants viewed persuasive charts with different sets of guardrails and reported trust, estimated rank in the dataset, expressed their perceived completeness of context, as well as subjective preference for different tasks. Across scenarios, guardrails improved trust, accuracy of performance judgments, and perceived completeness of context compared to the control. Taken together, the study offers practical guardrail sampling methods, evidence of their contextual benefits, and insights into participants' preferences.


\end{abstract}  
\section{Introduction}

Modern interactive time-series explorers make it easy for both novice and expert users to produce a factual, persuasive chart in seconds, before captioning it and sending it off over text or social media. Public health-related data exploration platforms, such as the Our World in Data COVID-19 Explorer~\cite{ourworldindata_covid-19_2020} and the Johns Hopkins dashboards~\cite{johnshopkinsuniversitycenterforsystemsscienceandengineering_covid-19_2020}, offer powerful controls for date windows, smoothing, and metric selection to inform users' health decisions. Finance platforms such as TradingView~\cite{tradingviewinc._tradingview_2025}, Yahoo Finance~\cite{yahooinc._yahoo_2025}, and Google Finance~\cite{googlellc_google_2025} similarly provide range, overlays, and comparators for easy customization of views pertaining to personal finance choices. These platform affordances are essential for domain-specific analyses for both experts and lay people alike, but they also make it easy to produce and disseminate misleading charts~\cite{lisnic_misleading_2023}. The same affordances that allow users to zoom in and focus on data points of their interest when the interface is used for exploration purposes may also lead to intentional or unintentional cherry-picking that may mislead both the user and their audiences~\cite{lisnic_visualization_2025}.

Deceptive visualizations and, consequently, visual misinformation are a pressing topic that has been attracting a growing body of work in the visualization community~\cite{lee_viral_2021,lisnic_misleading_2023, lo_misinformed_2022, pandey_how_2015}. There are many tools at visualization authors' disposal that assist in crafting misleading visualizations: framing and narrative choices in visualization expressed via titles, captions, selected time windows, axis ranges, and curated comparators may strongly shift reader interpretation and judgment, even when the underlying data are held constant~\cite{kong_frames_2018, kim_understanding_2021,lisnic_misleading_2023}. Visualizations are also persuasive~\cite{pandey_persuasive_2014} and are commonly shared on social media. Once they go viral, ``belief echoes''---residual attitudinal effects after correction---can persist even after misinformation is clearly retracted~\cite{thorson_belief_2016}. This reveals a pressing need to \textbf{design and deploy in-chart preemptive interventions that meet readers at the point of viewing and interpretation rather than relying solely on downstream fact-checking}.

We initiated the work to design such preemptive interventions by previously introducing \emph{visualization guardrails}: auxiliary context embedded directly into data explorer tools so that a \textit{focal item} (the item a user has chosen to display) cannot be read or shared in isolation~\cite{lisnic_visualization_2025}. In our original evaluation of guardrail designs, we found that simple guardrails that match the visual language of the main data successfully encourage skepticism while not interfering with the main chart~\cite{lisnic_visualization_2025}. However, the question of \textit{what} data to show and how to effectively sample it across data domains remains. For example, when exploring COVID-19 cases, it may matter which other countries are shown as guardrails: when a user is trying to understand Sweden's COVID-19 response, showing similar countries in geographic proximity might make sense. When exploring the stock price of General Motors, showing exemplars at the 10th, 50th, and 90th percentiles may help situate the data against the larger market. This paper aims to fill this gap and take guardrails from design space to practice. 

To identify which data is best displayed as a guardrail relative to a single \textit{focal item}, we selected five strategies from a larger set of options we considered, also illustrated in Figure~\ref{fig:teaser}: (1) \Percentiles{}---markers representing the percentiles of the dataset; (2) \PercentileExemplars{}---example items that are close to statistical percentiles; (3) \Clusters{}---example items that represent a cluster of the data; (4) \SemanticSim{}---items that are similar to the primary item based on external world knowledge about the data; (5) \RandomExemplars{}---a randomly-drawn subset of data, which we use as a naive baseline. In our experiments, we compare these to each other and to a \Control{} control condition (6).
As these guardrails are intended to be implemented by data exploration platforms, they span a spectrum with respect to the development effort and metadata needs: they range from simple statistical options to context-informed options. However, all of these methods can be precomputed as a new dataset is loaded and are feasible to integrate into existing explorers with minimal to no supervision.

To test the effectiveness of the sampling methods, we ran a \href{https://osf.io/cj8gt/overview}{preregistered}, mixed-design crowd-sourced study ($n=500$) comparing the five methods across two domain scenarios (COVID-19 and Stocks). We organize our evaluation around the following main research question: How do different guardrail sampling methods compare in effectiveness?
More specifically, we evaluated four outcomes: \textbf{Trust}---perceived chart trustworthiness, \textbf{Accuracy}---rank judgment of the focal item in the dataset, \textbf{Context}---a judgment of the sufficiency of context, and lastly \textbf{Preference}---participants' personal preference of a guardrail method selected from a lineup. 

Our findings indicate that, overall, \textbf{guardrails consistently outperform the \Control{} control across all measures}, with each having its own advantages.
Methods that surface statistical cues---\PercentileExemplars{}, \Percentiles{} and \Clusters{} work best when the goal is accurately judging the overall performance of the focal item. \SemanticSim{}, which shows real-world peers of the focal item, is especially effective when the goal is chart credibility. Meanwhile, \RandomExemplars{} is a surprisingly effective lightweight baseline that outperformed our expectations. In summary, our work provides an empirically validated roadmap for automatically contextualizing time series visualizations, quantifying the design trade-offs between guardrail strategies to reliably optimize user needs for either trust or accuracy.

\section{Background \& Related Work}

In this section, we first discuss interventions aimed at mitigating misleading charts at the times of creation, exploration, and consumption stages. We then explore how interactive explorers may enable cherry-picking. Finally, we review the \emph{guardrails} concept and design space that we build upon in this work.

\subsection{Interventions for Mitigating Misinterpretation}

There are several key issues that contribute to visualizations misleading their audiences, each requiring its own interventions. One is visual tricks: choices made while constructing charts, such as truncated axes, omitted baselines, and zoomed-in ranges, may lead their viewers to incomplete or biased interpretations~\cite{pandey_how_2015}. Researchers developed checkers that help authors mitigate risky specifications at creation time, before charts are published. Draco, for example, encodes design rules as constraints and automatically surfaces specification issues, recommending safer alternatives~\cite{moritz_formalizing_2019}. Practical visualization linting prototypes similarly flag command pitfalls programmatically and surface them to the author~\cite{mcnutt_linting_2018, hopkins_visualint_2020}. To help the audiences combat visual tricks that the authors did not catch (or intended to spread), literature proposed tools that present a check view or add salient annotations that can help viewers notice exaggerations without leaving the primary figure~\cite{fan_annotating_2022, ritchie_lie_2019}. More recently, researchers developed AI-assisted chatbots that transform views and describe the tricks upon user demand~\cite{das_misvisfix_2025}.

Another issue that leads to misleading visualizations arises during exploratory visual analyses, when reviewing many subgroups increases the likelihood of spurious patterns, a phenomenon known as the multiple comparisons problem~\cite{zgraggen_investigating_2018, zhao_controlling_2017}. Interventions addressing this problem propose incorporating safeguards that monitor user actions to validate whether the discovered pattern is statistically sound~\cite{zhao_controlling_2017}.

All interventions against misleading visualizations, however, share a common limitation in that they are not effective unless the viewer is skeptical in the first place and trusts the intervention. Once a misleading chart has been flagged as such, after it has already spread, it becomes an uphill battle for post-facto interventions, such as issuing warnings, conducting fact checks, and adding appended notes. Moreover, people often continue to rely on initial implications even after a clear correction has been published, known as the continued-influence effect; ‘belief echoes’ can persist long after a claim has been refuted or retracted~\cite{lewandowsky_misinformation_2012, thorson_belief_2016}. Platform dynamics compound this issue, as false or misleading items tend to proliferate faster than corrections, which widens the gap between the initial exposure and any subsequent correction~\cite{vosoughi_spread_2018}. Additionally, cognitive biases and motivated reasoning may lead audiences to not be convinced even by a corrected chart. Evidence shows that misleading visualizations can attract substantial engagement and rebuttal, yet the first impression remains mostly intact~\cite{lisnic_yeah_2024}.

\subsection{Persuasive Framing and Cherry-picking in Visualization}

We use the term \emph{persuasive visualization} to refer to charts presented in a communicative context to influence beliefs or decisions~\cite{chih_persuasive_2008, markant_when_2023}. One particularly common pathway to create misleading visualizations is cherry-picking items in data explorer platforms, where a user can use the platform to make a data point appear unusually good or bad and then share such a view on social media~\cite{lisnic_misleading_2023}. Contemporary data explorers, such as public health dashboards and retail investing platforms, expose various degrees of freedom for the user to adjust the view: selecting a suitable time frame, modifying axes and scales, and curating comparators. This freedom of modification is crucial for exploration but can also enable cherry-picking~\cite{lisnic_visualization_2025}. Prior work confirms that visual exploration choices can introduce selection bias, which can be exploited by adversarial framing and persuasive tactics~\cite{zhao_controlling_2017}. Literature also documents concrete deceptive tactics that were used to create charts that spread misinformation about the COVID-19 pandemic and vaccination measures, such as including curated subsets and comparators, as well as captioning and annotation~\cite{lisnic_misleading_2023}.

Whether deceptive or correct, harmful or useful, visualizations are often used for persuasion. Design and presentation choices can affect attitudes even when the underlying data remains unchanged. Prior work on the \emph{persuasive phase} of the chart lifecycle argues that once a view exits exploratory analysis, it enters a communicative context where framing and emphasis aim to persuade, rather than to discover~\cite{chih_persuasive_2008}. Furthermore, charts can be more persuasive than equivalent text summaries~\cite{stokes_striking_2023}. Presentation details, such as sources and annotations, can modulate perceived credibility and influence~\cite{pandey_persuasive_2014}. Persuasive impact also depends on viewers' prior beliefs and goals, which can shape what they extract from a visualization~\cite{markant_when_2023}. In modern data explorers, these persuasive tools are readily available to users through item selection, axis adjustments, and metric choices, among other features.

In short, persuasive use of charts and the widespread availability of cherry-picking affordances create the perfect conditions for the spread of misleading, yet ``data-driven'' messages, presumably against the intentions of the creators of the data exploration platforms that enable them. This motivates the interventions that add context at the moment of viewing or exploring.

\subsection{Guardrails: Concept and Design Space}

In our prior work, we introduced \emph{visualization guardrails} to combat cherry-picking in interactive explorers at both exploration and explanation times~\cite{lisnic_visualization_2025}. Visualization guardrails are an in-chart intervention that embeds contextual comparisons directly into an interactive data explorer. In our evaluation, we compared several presentation styles (displayed in the appendix, Figure~\ref{fig:old-guardrails}) and revealed a clear pattern: contextual items that share the same visual language with the main items (such as simple grayed-out lines) successfully revealed cherry-picking, increased skepticism, and were easy to understand. At the same time, more complex distributional and statistical visual designs were ineffective, presumably because they are more difficult to understand. Our prior work, however, focused solely on presentation styles and did not explore how to best select items to display as the guardrail.

\textbf{In this paper, we build on this evidence, focusing only on the proven effective type of guardrails, and ask the practical question our prior work left open: Which contextual lines should the system show}? We translate the design concept of guardrails into deployable sampling strategies that real explorers can precompute and render quickly. We formalize these sampling methods, analyze their effects across a series of tasks, and provide guidance for selecting a method to match platform and task goals.

\section{Guardrail Selection Methods}
\label{sec:guardrails}

Next, we first describe our approach for designing a variety of guardrail selection methods and then introduce each method implemented and evaluated in this work.

\subsection{Design Process}

One of the primary design goals was to generate a small set of implementable and conceptually-distinct guardrail selection strategies that we could evaluate both by logical analysis and with a study. Initially, all of the authors met to brainstorm ideas for selection strategies, prioritizing coming up with as many different ideas as possible. Our session was informed by the results of our prior work~\cite{lisnic_visualization_2025}, which showed that visually and conceptually simpler guardrails are more understandable and effective. We came up with a set of 11 distinct strategies, which we implemented and iteratively tested out on real-world data. We then conducted multiple rounds of (i) adjusting the conceptual and implementation details to make sure that guardrails result in understandable, effective, and uncluttered views, and (ii) grouping together or further differentiating similar ideas, eventually paring down to five distinct sampling strategies.

Another key choice we had to make is \emph{how many} comparators (exemplars, statistical markers) to show. Designers have to find a balance between a good coverage of the data space and clutter. Prior research on showing COVID-19 forecasts~\cite{padilla_multiple_2022} suggests that trust hits a plateau when showing 6--9 forecasts, while prediction accuracy plateaus at 5--7 forecasts. We found that when considering historical data associated with different items, as opposed to modeled forecast data, there is a lot of volatility in the data (e.g., prices of different stocks fluctuate independently of each other), leading to more clutter. As such, we believe that displaying $n=5$ contextual items alongside the focal item is a justifiable choice, yet we note that our methods work with fewer or more items as well.

\subsection{Selection Methods}

As a result of our design process, we developed five selection methods for evaluation, each with its own advantages and potential pitfalls. Figure~\ref{fig:teaser} summarizes the five methods, spanning data-driven, world-knowledge, distributional, and exemplar-based comparators. One method, \RandomExemplars{}, serves as a naive baseline against which we evaluate the rest, termed Context-rich guardrails. \textit{Context-rich guardrails} use either the dataset or world knowledge to select useful contextual items. Most guardrails are \textit{exemplar-based}, meaning they display other items from the dataset and thus are of the same data type as the focal item. One method is \textit{abstract mark-based}; it shows statistical information and does not correspond to any actual item from the data.

\paragraph{\RandomExemplars{}}
The \RandomExemplars{} condition shows $n$ items randomly drawn from the same dataset alongside the chosen focal items, without regard for equal distribution or meaning. We primarily implemented this condition as a naive baseline of a guardrail. This method's main strength lies in its simplicity and applicability to any domain and dataset. Additionally, even small random samples can sufficiently convey distributional cues, such as noisiness vs homogeneity~\cite{hullman_hypothetical_2015,albers_task-driven_2014}. At the same time, the randomness may also lead to uninformative or outright misleading comparisons.

\paragraph{\Percentiles{}}
A more data-informed (yet still domain-agnostic) approach would be to show distributional information about the rest of the dataset alongside the focal item. \Percentiles{} shows aggregate percentile bands. We calculated five roughly equally spaced percentiles---5th, 25th, 50th, 75th, and 95th---and show them as lines alongside the focal item. 
We create the percentile lines by calculating the five percentiles at each timestep $t$ across the entire dataset. So, in essence, these bands do not necessarily correspond to ``real'' data or trends found in the dataset and serve as synthetic distributional markers.

\paragraph{\PercentileExemplars{}}
In order to still convey the distribution but anchor it in real items and thus potentially increase understandability, we designed the \PercentileExemplars{} strategy. Here, for each previously calculated synthetic percentile line, we find the single item (stock or country) that most closely tracks the percentile line using the minimum least-squares distance across all time steps.
This method allows for apples-to-apples comparison with the focal item while still conveying the distribution. Alongside the item labels, we additionally display the percentile that they serve to represent.

\paragraph{\Clusters{}}
The \Clusters{} condition similarly aims to display the variety of trends in the dataset while anchoring it to actual items. In this strategy, we first cluster all time series in the dataset into $n$ clusters using the $k$-means algorithm. For each cluster, we find the item closest to the cluster centroid by computing the minimum L2 distance.
The advantage of this method compared to \PercentileExemplars{} is that it does not prioritize uniformly-trending items that best align with a synthetic percentile curve. In cases where a dataset has a lot of items that show big increases or big drops throughout the period, clustering such shapes would allow us to surface them as context and potentially get a more realistic view of the variation of trajectories in the data. On the other hand, this condition is somewhat less intuitive to explain to the user.

\paragraph{\SemanticSim{}}
One limitation of all of the above methods is that they do not capture any of the potentially meaningful semantics of the data. For instance, when comparing COVID-19 case curves it might be most useful to look at countries from the same region, with similar demographics, alongside other metrics. The goal of this method is to sample a set of $n$ semantically comparable peer items that would make sense to contextualize the focal item in real-world scenarios. This could be the same sector and similar market cap for stocks, or the same region and similar demographics for countries.
Our implementation employs a large language model (LLM) ensemble and a high-agreement filtering mechanism to select robust comparators. We generate ten independent samples by querying the model (ChatGPT-5) with a fixed prompt that requests the five most appropriate comparators for the focal item and task. We then apply a majority-vote consensus heuristic, retaining any entity that achieves a high level of inter-response agreement (appearing in seven or more of the ten generated samples). The full prompts are detailed in the appendix.

There are a number of alternative methods that could be employed here and that we considered in the process: collecting domain-specific metadata (such as UN region classes for countries) or collaborating with a subject matter expert to identify an appropriate context for each focus item. The LLM approach we used offers significant flexibility as it works for many domains, especially those of interest to data explorers targeting the general public. It also offers low overhead; however, we urge any implementation to first thoroughly investigate whether the method yields appropriate results.

\begin{figure*}[htb]
    \centering
    \begin{subfigure}[t]{0.45\textwidth}
        \centering
        \includegraphics[width=\linewidth]{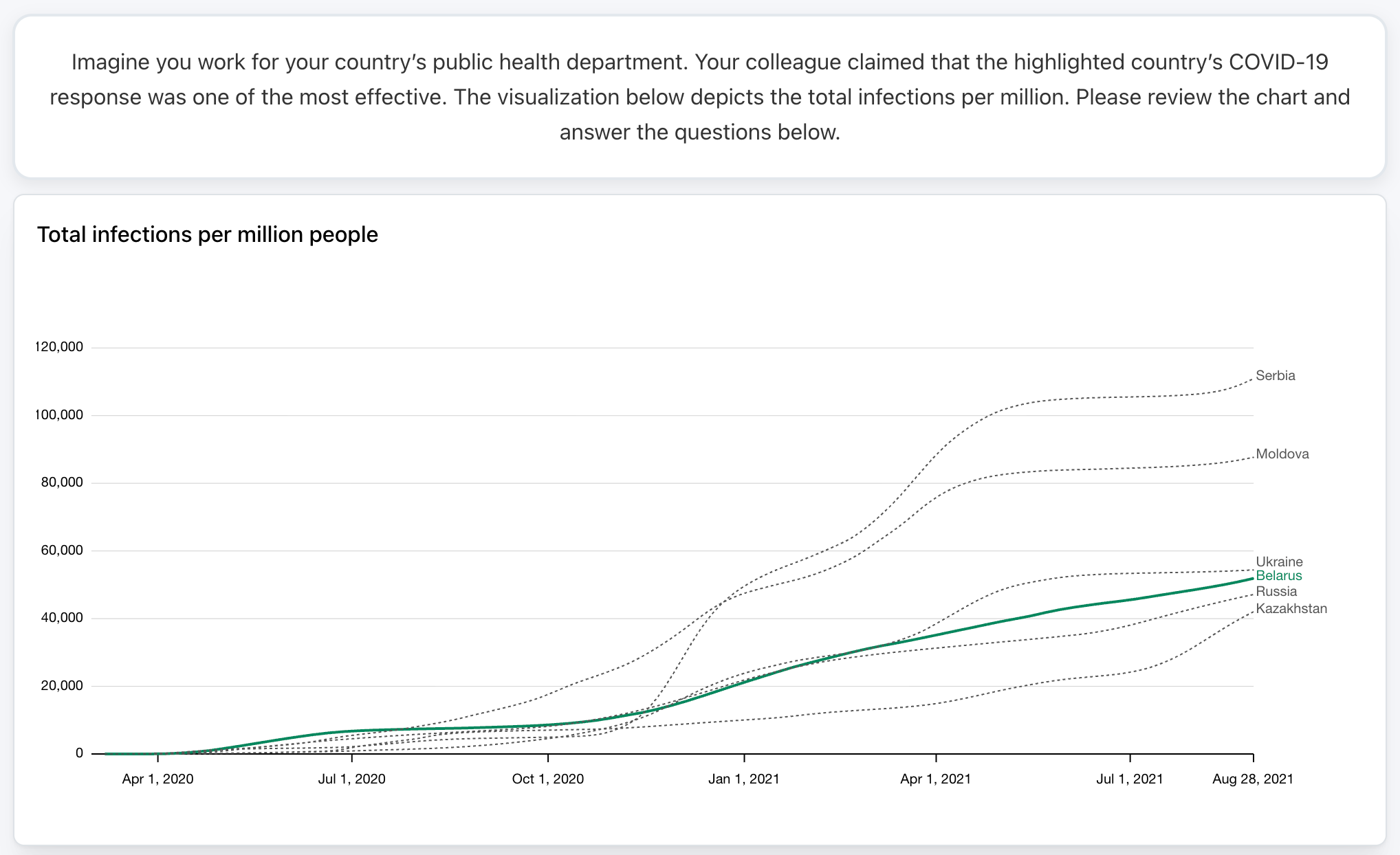}
        \caption{\SemanticSim{} in the COVID scenario}
        \label{fig:screenshot_ss}
    \end{subfigure}
    \hfill
    \begin{subfigure}[t]{0.45\textwidth}
        \centering
        \includegraphics[width=\linewidth]{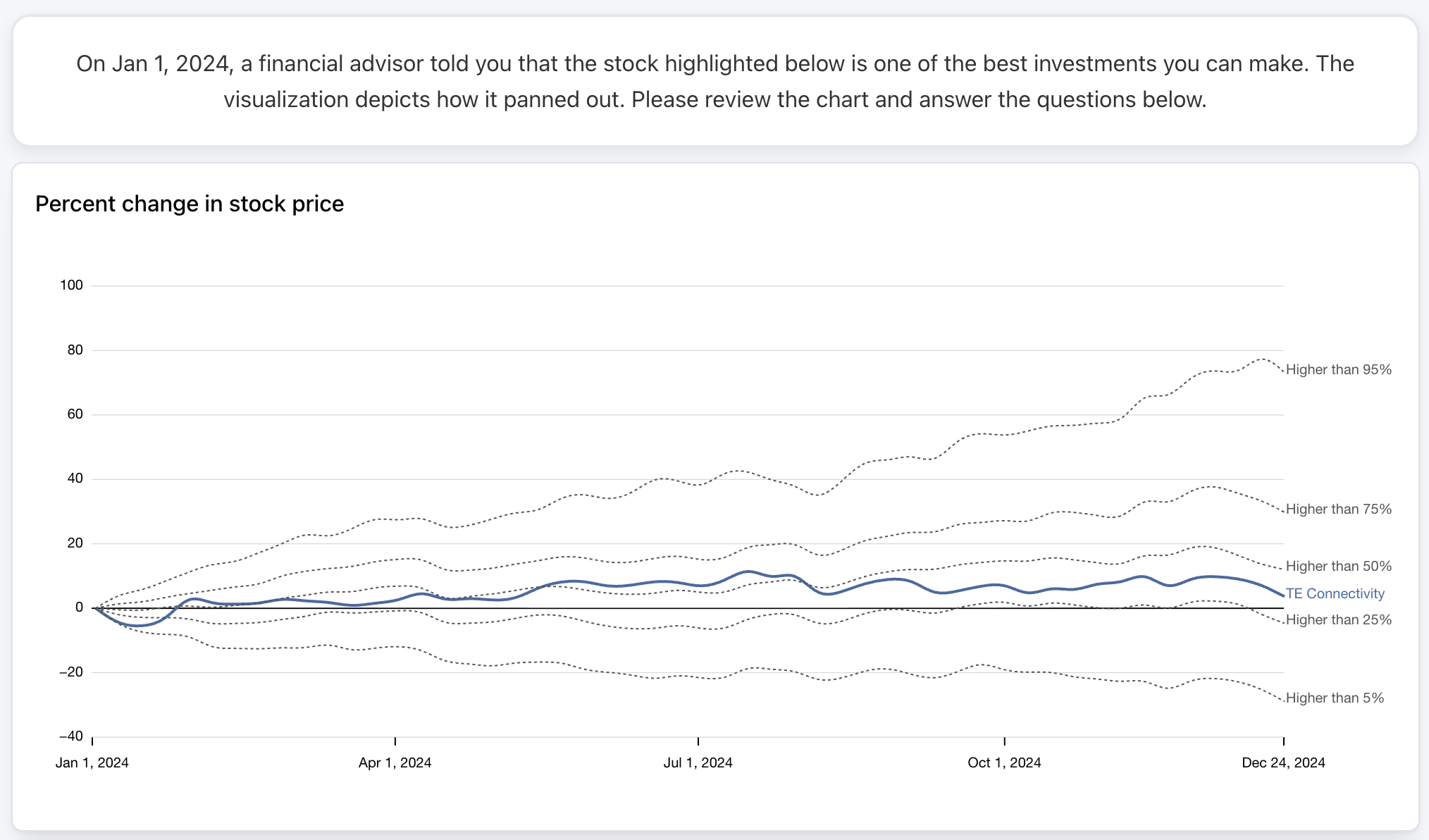}
        \caption{\PercentileExemplars{} in the Stock Scenario}
        \label{fig:screenshot_pe}
    \end{subfigure}

    \vspace{-0.2cm}
    \caption{Examples of our guardrails as seen in the Task 1 stimulus. Figure (a) illustrates the \SemanticSim{} strategy, showing items semantically similar to the focal entity (here, countries demographically and geographically similar to Belarus). Figure (b) illustrates the \PercentileExemplars{} strategy, which instead selects context based on their percentile bins in the global distribution.}
    \vspace{-0.5cm}
    \label{fig:combined_screenshots} 
\end{figure*}

\section{Evaluation Methodology}

To identify user preferences across guardrails sampling methods, as proposed in Section~\ref{sec:guardrails}, we conducted a \href{https://osf.io/cj8gt/overview}{preregistered} mixed-design crowd-sourced study. In this section, we outline the recruitment and assignment of participants, study design, and analysis methodology. All prompts, stimuli, datasets, analysis code, and other materials are provided in the supplemental materials. Our study, developed with reVISit~\cite{cutler_revisit_2026}, can be reviewed by following the \href{https://vdl.sci.utah.edu/guardrail-samples-study/}{link}. The code for the study is available on \href{https://github.com/visdesignlab/guardrail-samples-study}{GitHub}.

\subsection{Study Procedure}

In designing our study, we aimed to elicit guardrail preferences across a variety of measures, task framings, and scenarios. As such, our study utilized two datasets, two task types, and several measures of preference, described in more detail below. The factors for data scenario (COVID-19 vs. Stock Performance) and the specific guardrail conditions (only two of the five guardrail types were shown to an individual participant) were manipulated between subjects. The factors for the focal item (three specific items seen by all participants) and the presence of a guardrail versus the control condition were manipulated within subjects. This approach allowed us to compare the effect of having a guardrail versus no guardrail and account for variation across specific focal items at the individual participant level.

In terms of process, each participant was assigned to a scenario---either COVID-19 cases by country or stock performance over time---and completed two tasks, followed by a survey. Due to a combination of randomization and participant dropout, the conditions are not perfectly balanced. Additionally, due to technical issues, we separately followed up with participants to complete the post-study survey for an additional payment of \$0.25 and received 468 responses. In the survey, we collected their self-reported familiarity with charts, statistics, the domain of the scenario, as well as their political leaning.  We report the results of the post-task survey in the supplemental materials; however, we note that the insights did not meaningfully inform the results of our main analysis, and we thus omit them from the discussions below.

\subsubsection{Task 1: Chart \& Survey}

The first task was \textbf{Chart \& Survey}, in which we presented the participants with a chart accompanied by a persuasive caption. In the COVID scenario, the chart was shared by a colleague in public health who was advocating for adopting a given country's pandemic protocol, describing it as ``one of the most effective'' (seen in Figure~\ref{fig:screenshot_ss}). In the Stocks scenario, the chart was shared by a financial advisor who had promoted a given investment, by similarly describing it as ``one of the best investments'' (seen in Figure~\ref{fig:screenshot_pe}). Each participant saw three such charts (with different data) in random order, one by one: one Control chart (with no guardrails whatsoever), and two randomly selected guardrail methods. In all cases, however, the focal item was selected to be middle-of-the-pack as to evaluate the usefulness of providing guardrails as context. The specifics of data selection are discussed further in Section~\ref{sec:data}.

For each chart, participants rated how trustworthy they found the chart on a 7-point Likert scale, estimated the focal item's position in the entire dataset from 0 to 100 (equivalent to percentile rank), and lastly reported how appropriate they found the context in the chart to make an informed decision given the scenario, also on a 7-point Likert scale. After seeing the three charts, participants completed a simple attention check question, which can be seen  \href{https://vdl.sci.utah.edu/guardrail-samples-study/stage-1/MTdrZWtXYTNtR1hvUU9zNit1Z1pRQT09}{here}.

\subsubsection{Task 2: Preference Selection}

The second task was \textbf{Preference Selection}, in which the participants saw a lineup of four charts. Each chart depicted the same focal country or stock, but with different guardrails: \RandomExemplars{}, \Clusters{}, \PercentileExemplars{}, and \SemanticSim{}. We excluded the \Percentiles{} here to limit the conditions to Exemplar-based guardrails.

We asked the participants to select the chart that they find the most useful, and then provide their rationale in an open text response field. To assess the differences in guardrail preferences according to the \textbf{specifics of the task}, we randomly assigned the participants into two groups that differed in their task framing. The first group was asked to complete a task in what we called a \textbf{Holistic} framing: they were asked to choose a chart that best supports a relative evaluation. For instance, in the COVID condition, the participants were asked to judge the effectiveness of a country's COVID response. Such a task, in theory, is best judged when compared to demographically and geographically similar countries; thus, we hypothesized that most would select \SemanticSim{}. The second group saw what we called a \textbf{Precise} framing, which tasked the participants to choose a chart that best supports an overall or global performance. For example, in the context of stocks, they were asked to evaluate the return of the stock. This task, in contrast, is best judged on an absolute scale (i.e., when choosing an investment to maximize profit, how peer companies did in the market does not matter as much as all alternative investments), thus we theorized that most participants would select the \PercentileExemplars{} view.

\subsection{Procedure \& Recruitment}

We first conducted an informal pre-pilot with colleagues from our university. As a result, we clarified task wording and introduced the two diverging framings in the Preference Selection task, as feedback indicated that participants would often assume one of these two task framings. We then conducted a pilot study with 85 crowd-sourced participants on Prolific. We used the results of the pilot to estimate the payment based on completion time, as well as to size our experimental sample using a power analysis. We aimed to detect moderately-sized effects based on our pilot results with 80\% power and at the standard statistical significance level of 5\%. As a result, we recruited 500 participants for the main study. The median completion time was just under 7 minutes, and we paid each participant \$2.10 (for an hourly rate of \$18/hour).
Our IRB deemed the study exempt from full review. All participants gave informed consent, and we collected no identifying data.

\subsection{Dataset, Stimuli and Framing Materials}
\label{sec:data}

We sourced the COVID-19 dataset from Our World in Data~\cite{ourworldindata_covid-19_2020}. The charts in the study showed the participants cumulative infections per million inhabitants from April 2020 to August 2021. 
We used cumulative infections as opposed to the daily rate because our pre-pilots showed that daily variance made the charts visually noisy and less clear for the judgment tasks. We collected the S\&P 500 stock prices via a script using Yahoo Finance API and showed data for the calendar year 2024~\cite{yahooinc._yahoo_2025}. To reduce noise and smooth the trajectories for clearer presentation, similar to the COVID-19 data, we plotted the weekly closing prices and transformed the values into percentage changes over the period, all starting at 0 at the beginning of the chart. This is consistent with common stock market data exploration sites, such as Yahoo Finance and Google Finance.

We designed the charts---seen in Figure~\ref{fig:combined_screenshots}---to be consistent across conditions and aesthetically minimalist as to not interfere with performance. Our initial work on evaluating guardrail designs showed that it is important to clearly visually distinguish the guardrails from the main data~\cite{lisnic_visualization_2025}. Consequently, we designed each chart to highlight the focal item with a saturated color and a slightly thicker stroke, while auxiliary guardrail lines were grayed and dashed, allowing them to recede into the background while still remaining salient.

In choosing the focal items to highlight, we aimed to select the ``(lower) middle of the pack'' views that could believably be used for persuasive visualization (thus not the worst or blatantly unconvincing), yet underperforming enough that contextual data would provide revealing evidence. In other words, we wanted to ensure that the decision to trust or distrust the chart is not trivial and be able to measure the relative effect of contextual data on persuasiveness.
To select such datapoints for Stocks scenario, we filtered the set of S\&P 500 companies to those that met the following three criteria: (i) roughly visually smooth weekly trajectories, to control for potential volatility interference; (ii) no significant dips below 0\% to avoid significant floor cues; and (iii) approximately 35th percentile performance for the year across the entire dataset, so not the best but also not the worst. We selected three stocks closest to the 35th percentile that satisfied these criteria for task 1: TEL (TE Connectivity), COR (Cencora), and CHD (Church \& Dwight), and one for task 2: VZ (Verizon). To select countries to show in the COVID scenario, we followed a similar protocol, but in this case, sampling from the 65th percentile of cases per capita, as with COVID the ``goodness'' metric is flipped and lower is better. We again selected the three closest items that satisfied the criteria for task 1, yielding Greece, Germany, and Belarus, and one for task 2: Norway. 

\subsection{Analysis Overview}

We adhered to our preregistration, which can be found \href{https://osf.io/cj8gt/overview}{here} and describes the analysis in detail. We excluded responses that: (i) failed attention checks, (ii) open-ended rationales that were non-substantive or gibberish, (iii) incomplete trials, or (iv) data missing because of technical issues. As a result, we excluded 13 participants and analyzed 487 responses. The raw responses, analysis scripts, and participants' anonymized demographic data are available in the supplemental materials.

For Task 1 (Chart \& Survey) outcomes (Trust, Accuracy, and Context), we investigated two main hypotheses (see Table~\ref{tab:hypotheses}). First (H1), we tested if adding any guardrail improved responses compared to the baseline \Control{} condition. Second (H2), we tested if Context-rich guardrails performed better than a naive \RandomExemplars{} guardrail. We employed linear mixed-effects analyses, including participant and stimulus as random intercepts. Fixed effects were evaluated using likelihood-ratio tests, followed by planned pairwise contrasts with Holm correction.

For Task 2 (Preference Selection) responses, we tested which selection method was preferred most often across two task prompts to identify whether users' preferences depend on the question at hand (H3, see Table~\ref{tab:hypotheses}). We first used a chi-square test of independence to determine if the prompt type (Holistic vs.\ Precise) influenced the distribution of chosen chart types. Within each prompt, we ran preregistered one-sided two-proportion z-tests, comparing the hypothesized winner to the other options. We adjusted $p$-values using Holm correction, and we report proportions with Wilson 95\% Confidence Intervals (CIs) and Cohen’s $h$. 

Finally, we performed a thematic analysis with inductive open coding on the Task 2 preference rationales. To code the results, the first author conducted an initial pass over the data, assigning individual codes to participants' preference rationales. The last author then completed their own pass, suggesting code edits, deletions, and combinations. The first and last authors met to further group codes into higher-level categories and organize the final codebook.

\begin{table*}[t]
\centering
\small
\setlength{\tabcolsep}{5pt}
\renewcommand{\arraystretch}{1} %
\caption{Overall, guardrails improved \emph{Trust}, \emph{Context}, and relative-performance \emph{Accuracy} versus Control (H1a-c supported). Context-rich guardrails did not exceed a Random guardrail on Trust or Context (H2a, H2c not supported), but did improve Accuracy relative to Random (H2b supported). Task framing (Holistic vs.\ Precise) did not alter guardrail preferences (H3a-c not supported).}
\vspace{-0.3cm}

\label{tab:hypotheses}
\begin{tabular}{p{0.015\textwidth}p{0.40\textwidth}p{0.10\textwidth}p{0.4\textwidth}}
\toprule
\textbf{ID} & \textbf{Statement} & \textbf{Support} & \textbf{Findings summary} \\
\midrule
\multicolumn{4}{l}{\textbf{H1: Guardrails vs No Guardrails}}\\
\midrule
H1a & Guardrails increase \emph{Trust}. & \textbf{Supported} & Higher with any guardrail vs \Control{} with a small but reliable effect, about +0.5 Likert points. \\
H1b & Guardrails improve \emph{Accuracy}. & \textbf{Supported} & Significantly lower error for all guardrails except \SemanticSim{}, where the results are highly variable. \\
H1c & Guardrails increase \emph{Context}. & \textbf{Supported} & Significant improvement of about +1.5 Likert points, on average, compared to \Control{}. \\
\midrule
\multicolumn{4}{l}{\textbf{H2: Context-rich Guardrails vs Random Items}}\\
\midrule
H2a & Context-rich guardrails increase \emph{Trust} over \RandomExemplars. & \textbf{Not supported} & No improvement over random, with the exception of \SemanticSim{} in the COVID scenario. \\
H2b & Context-rich guardrails improve \emph{Accuracy} over \RandomExemplars. & \textbf{Supported} & Significantly lower error for all guardrails except \SemanticSim{}, where the results are highly variable. \\
H2c & Context-rich guardrails increase \emph{Context} over \RandomExemplars. & \textbf{Not supported} & No improvement over random, with the exception of \SemanticSim{} in the COVID scenario. \\
\midrule
\multicolumn{4}{l}{\textbf{H3: Task Framing Influence on Guardrail Preferences}}\\
\midrule
H3a & Chosen chart \emph{Type} differs between \emph{Holistic} vs \emph{Precise} prompt. & \textbf{Not supported} & No significant effect of prompt on preference choices. \\
H3b & Under \emph{Holistic}, \SemanticSim{} is chosen most frequently. & \textbf{Not supported} & \SemanticSim{} not reliably preferred more under Holistic vs other types. \\
H3c & Under \emph{Precise}, \PercentileExemplars{} is chosen most frequently. & \textbf{Not supported} & \PercentileExemplars{} not reliably preferred more under Precise vs alternatives. \\
\bottomrule
\end{tabular}
\vspace{-0.5cm}
\end{table*}

\section{Study Results}

In this section, we report the results of our evaluation across the two tasks. Table~\ref{tab:hypotheses} presents a concise summary of the study results relative to the preregistered hypotheses. The exact $p$-values for all preregistered hypothesis tests are reported in Appendix Table~\ref{tab:exact_pvalues}.

\subsection{Task 1 Results}

\begin{figure*}
    \centering
    \includegraphics[width=\linewidth]{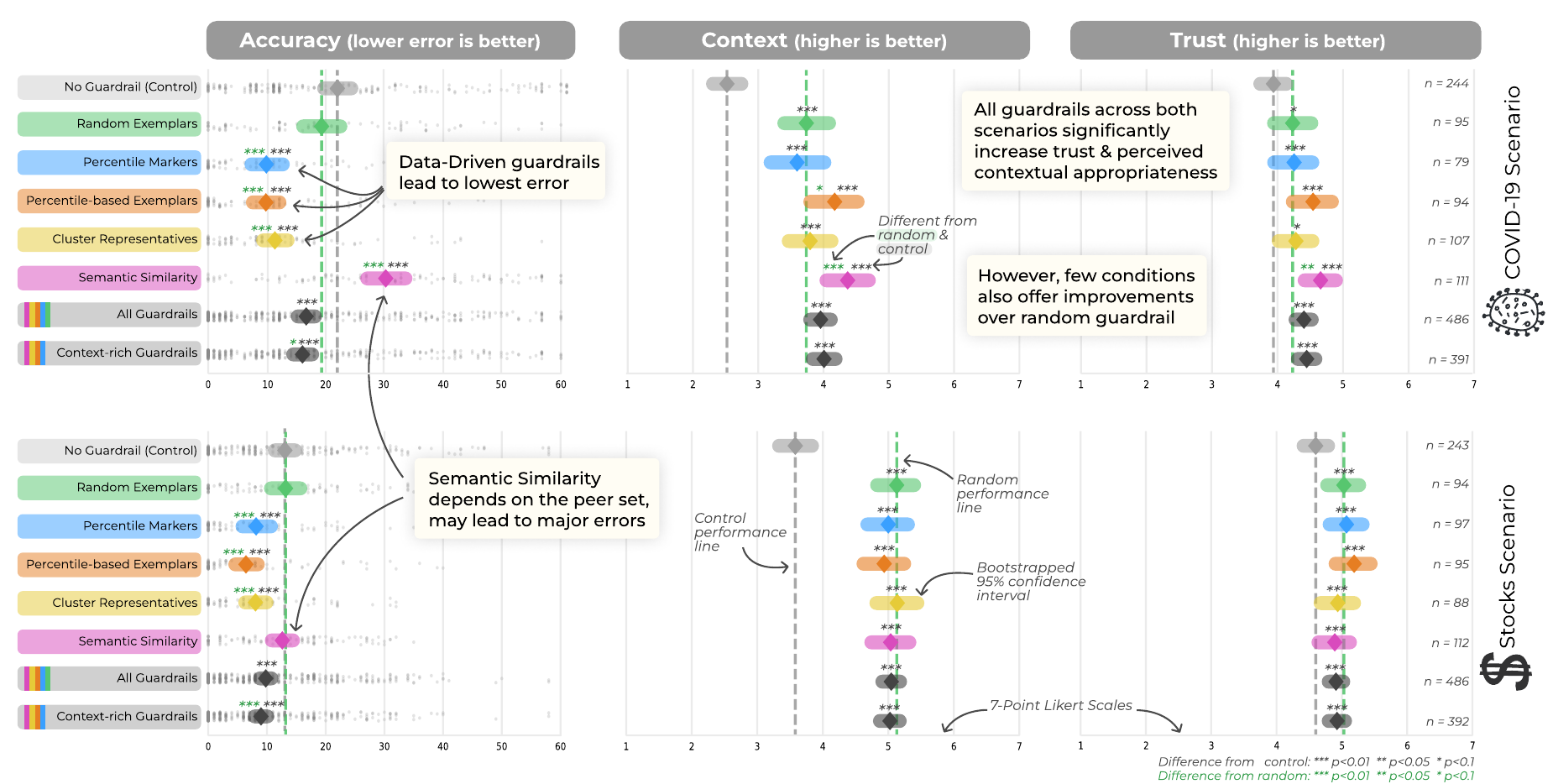}
    \vspace{-0.5cm}
    \caption{Across all tasks and scenarios, guardrails improve accuracy, trust, and context. For trust and context, \RandomExemplars{} performs as well as data-driven guardrails, and \SemanticSim{} performs better in the COVID scenario. Data-driven guardrails---\PercentileExemplars{}, \Percentiles{}, and \Clusters{}--- however, offer a significant improvement in accuracy. \SemanticSim{} method greatly depends on the distribution of the semantically-similar peers, thus may lead to major accuracy errors. Note that the statistical significance markers follow from mixed-effects models that also account for participant and item effects and may not directly reflect visible confidence interval differences.}
    \vspace{-0.5cm}
    \label{fig:task1_overview}
\end{figure*}

Figure~\ref{fig:task1_overview} presents an overview of the results of Task 1 (Chart \& Survey). Across scenarios and measures, we see that presenting \emph{any} type of guardrail improved responses, however the specifics vary.

We observe that respondents consistently report \textbf{higher trust in charts that do feature guardrails}, confirming hypothesis H1a (COVID: $\chi^2 = 16.46$, $p < 0.001$, stocks: $\chi^2 = 22.33$, $p<0.001$). The effect, albeit statistically significant, is not large: trust increases by roughly half of a Likert point with guardrails. Notably, while guardrails overall are associated with higher reported trust, using Context-rich guardrails does not yield a difference relative to \RandomExemplars~(COVID: $\chi^2 = 1.83$, $p = 0.18$, stocks: $\chi^2 = 0.324$, $p = 0.57$), leading us to reject hypothesis H2a. In other words, \textbf{any random guardrails are approximately as effective at increasing trust as the more meaningful samples.}

Participants' responses to the context appropriateness question (e.g., ``Do you feel that the visualization provided appropriate context to make an informed decision about your investment?'' for stocks) follow a similar pattern, however with stronger effects. \textbf{Guardrails increase the context substantially compared to \Control{}} (COVID: $\chi^2 = 100.17$, $p < 0.0001$, stocks: $\chi^2 = 113.08$, $p<0.0001$), by an average of 1.5 Likert points and moving the results from below to above the midpoint, confirming H1c. Here again, notably, \textbf{the \RandomExemplars{} condition performs as effectively} as the more Context-rich guardrail methods (COVID: $\chi^2 = 1.78$, $p = 0.18$, stocks: $\chi^2 = 0.35$, $p = 0.56$), rejecting H2c.

We also confirm that \textbf{guardrails significantly decrease the error when estimating where the data point falls in the distribution}, reducing it by an average of 5.3 points in COVID scenario, and 3.25 points in Stocks (COVID: $\chi^2 = 15.7$, $p<0.0001$, stocks: $\chi^2 = 12.03$, $p<0.001$) and confirming our hypothesis H1b. The Context-rich results also prevail in the Stocks scenario, decreasing the error on average by 4.2 more points compared to the \RandomExemplars{} guardrail ($\chi^2 = 11.18$, $p<0.001$). In the COVID scenario, however, the improvement was marginal ($\chi^2(1)=3.11$, $p=0.08$). We also note that, as shown in Fig.~\ref{fig:task1_overview}, all methods \emph{except} \SemanticSim{} significantly improved accuracy; \SemanticSim{} underperforms all other guardrails, including \Control{} in the COVID scenario. Our findings confirm a potential negative effect of this sampling strategy: while \SemanticSim{} might be trustworthy and provide helpful context, it may also misleadingly frame the data within a non-representative subgroup: the semantic subset may be significantly skewed relative to the overall distribution.

Lastly, we note that the results are largely consistent across the two data scenarios, with several deviations. Firstly, the results primarily vary in the \SemanticSim{} condition, which follows from the fact that the guardrail items are highly dependent on the specifics of the domain and the focal item and thus result in higher variance. Secondly, the absolute values across all three measures are notably different: participants find Stock charts more trustworthy and complete than COVID, and make more accurate performance judgments about them. This result likely follows from the fact that participants are both more familiar with (and more biased by) the ubiquitous COVID-19 visualizations.

\subsection{Task 2 Results}

\begin{figure*}[htb]
  \centering
  \includegraphics[width=\linewidth]{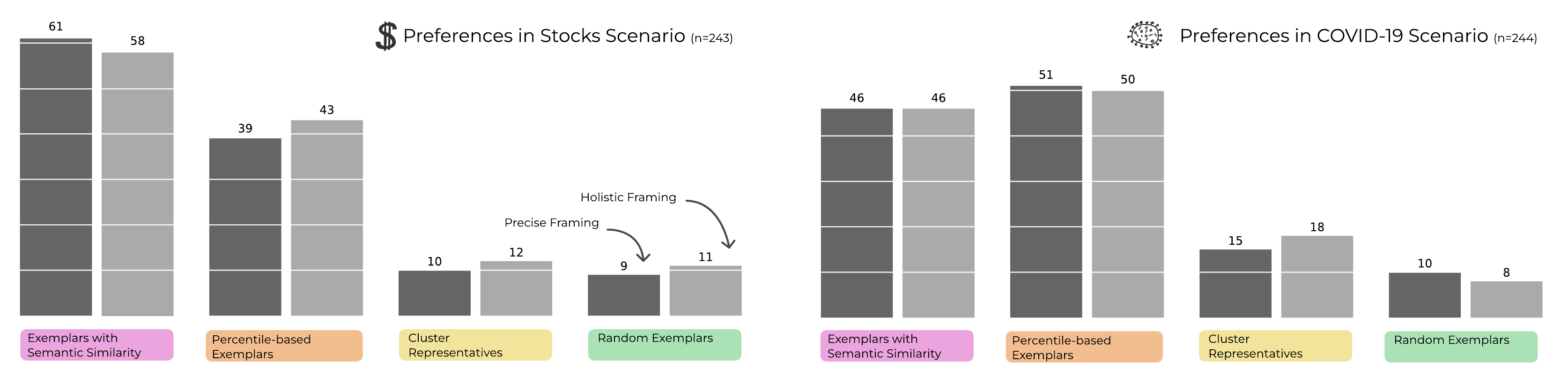}
  \vspace{-0.8cm}
  \caption{When asked to choose, participants prefer \SemanticSim~ and \PercentileExemplars~ at roughly the same rate, independent of framing, even though we considered \SemanticSim{} the better choice for the Holistic framing (the question targeted judgment relative to peers), and percentiles the better choice for Precise framing (the question targeted estimating absolute performance). \Clusters~  and \RandomExemplars~ are infrequently chosen for these tasks.}
  \label{fig:preferences}
\end{figure*}

Figure~\ref{fig:preferences} shows participants' responses in Task 2: Preference Selection. We observed no significant framing effect on preferences, rejecting our hypothesis H3a. The distribution of the selected methods did not differ between Holistic and Precise prompts in either scenario (COVID: $\chi^2 = 0.50$, $p = 0.92$, stocks: $\chi^2 = 0.55$, $p = 0.91$). Consequently, participants also did not choose the \SemanticSim{} guardrails more often in the Holistic prompt or the \PercentileExemplars{} in the Precise prompt, leading us to reject hypotheses H3b and H3c, respectively.

Interestingly, the results are consistent across both scenarios. Although not a tested hypothesis, we suspected that \SemanticSim{} would be preferred for COVID-19 data, as pandemic data was typically compared against peers, while \PercentileExemplars{} would be more preferable to evaluate stock data, since investors are typically not limited to investing among a group of stocks. Taken together, the \textbf{selection results primarily reflect participants' general preferences and do not vary with the specifics of the task prompt or the domain}. By and large, participants prefer either \SemanticSim{} or \PercentileExemplars{} guardrails to complete their tasks. In summary, although both sampling methods performed similarly in Task 1, Task 2 demonstrates a significant user inclination toward specific guardrail types.

\section{Qualitative Rationale}

\begin{figure*}[t]
\vspace{-0.3cm}
\centering
\small
\setlist[enumerate]{nosep}
\begin{minipage}{0.96\textwidth}
\begin{multicols}{2}
\begin{enumerate}[leftmargin=*, label=\textbf{\arabic*.}]
  \item \textbf{Context strategies}
    \begin{enumerate}[label*=\arabic*., leftmargin=1.2em]
      \item \textbf{Selection coverage}
        \begin{enumerate}[label*=\arabic*., leftmargin=1.4em]
          \item \texttt{Broader contextual coverage}
          \item \texttt{Broader coverage by distribution}
        \end{enumerate}
      \item \textbf{Similar contextual peers}
      \item \textbf{Specific entities included}
      \item \textbf{Information specificity}
        \begin{enumerate}[label*=\arabic*., leftmargin=1.4em]
          \item \texttt{More information preferred}
          \item \texttt{Objective numbers preferred}
        \end{enumerate}
    \end{enumerate}

  \item \textbf{Chart aesthetics preferences}
    \begin{enumerate}[label*=\arabic*., leftmargin=1.2em]
      \item \textbf{Declutter and clarify}
        \begin{enumerate}[label*=\arabic*., leftmargin=1.4em]
          \item \texttt{Clean chart}
          \item \texttt{Fewer overlapping lines}
        \end{enumerate}
      \item \textbf{Scale preferences}
        \begin{enumerate}[label*=\arabic*., leftmargin=1.4em]
          \item \texttt{More granular axis}
          \item \texttt{Wider axis range}
        \end{enumerate}
    \end{enumerate}

  \item \textbf{Perceived accuracy \& trust}

  \item \textbf{Focal series take-away}

\end{enumerate}
\end{multicols}
\end{minipage}
\caption{Final codebook used for open-ended responses. We organize the themes based on the highest level of codes. Each text response may be described by one or more codes. We excluded codes that indicated no preference or a clearly mistaken justification.}
\label{fig:codebook}
\vspace{-0.5cm}
\end{figure*}

To understand \emph{why} participants selected certain charts, we conducted an \emph{exploratory} analysis of open-ended text rationales. Figure~\ref{fig:codebook} shows our codebook used to annotate the responses. Below, we present three themes that describe the codebook.

\subsection{Theme 1: Preference for a Specific Context}

One way in which participants expressed their preference for a guardrail was by describing their desire for the specific property that the guardrail displayed. Codes 1.1--1.4 in the codebook reflect the preferences for contextual strategies that fairly directly map onto the specific guardrail selection methods we tested. For instance, some participants described preferring peers that make sense together in the real world (same sector for stocks; neighboring regions, similar demography for countries). As such, they chose the \SemanticSim{} method (code 1.2) and cited seeking ``apples-to-apples'' comparisons. One participant described preferring the chart that showed Norway's immediate neighbors: \hquote{I was influenced by how close the other countries’ data was geographically to Norway, because this closeness/similarity meant that you might be comparing similar countries, so getting rid of some of the factors influencing COVID-19 infections.}

At the same time, other participants had different expectations of best context. Participants who selected \PercentileExemplars{} charts (code 1.1.2) cited the preference for the ability to make global comparisons and glean the distribution: 
\hquote{This chart has percentages that seem to indicate how a labeled nation fares among the global COVID-19 outbreaks. It showcases nations with low outbreaks versus nations with high outbreaks in a simple manner.} Notably, these preferences reflect personal expectations, rather than a selection that best matches the task or scenario---as previously discussed in the quantitative results.

\subsection{Theme 2: Preference for Chart Aesthetics}

Moreover, the actual contextual content was not the only reason for participants' preferences. Many of the responses cited clean layouts, reduced overlap (codes 2.1), and axis scaling (codes 2.2) as rationales for their guardrail preferences: \hquote{The highest total of infections per million people in chart B only goes up to 80,000 and increments of 10,000 vs 20,000 in the other choices, showing slightly more detail in the changes over time.} While chart aesthetics and clarity of guardrails were not central to our guardrail design, they are often the primary reason for participants' preferences. This highlights an important challenge of competing demands: when attempting to show, for instance, semantically-relevant contextual data, one may end up with a more cluttered chart as a side effect. For instance, one respondent simply reported their preference being due to \hquote{there [being] less overlapping elements for the other countries}, emphasizing chart clarity above all.

\subsection{Theme 3: Preference for Specific Takeaways}

In another set of preferences, participants chose their preferred chart not because of the context itself, but rather based on the light in which it presented the focal item (codes 3, 4).
Text responses sometimes directly referenced the performance of the focal country or stock, using terms such as \hquote{middle of the pack}, \hquote{no big declines}, or \hquote{stable}. Participants often prefer a chart that reveals a specific quality of the target data that a condition shows, such as: \hquote{It shows it is middle of the pack, whereas others showed it was less so}. Alternatively, others simply prefer seeing the focal country performing better or worse, whichever aligns with their expectation: \hquote{It showed in comparison to others, there was no period of significant decline, and the overall performance was positive [...]}

\section{Discussion}

In this section, we interpret the findings and articulate their implications for visualization practice. We highlight the key trade-offs and limits that guide the implementations of guardrails in real systems.

\subsection{Guardrails help, but there’s no one-size-fits-all}

Across both the COVID and Stock scenarios, the presented guardrails consistently improved context and generally increased trust. Gains in accuracy, however, were concentrated in data-driven methods that explicitly encode distributional structure. This pattern implies that there is no \textit{single best} guardrail; instead, the suggested guardrail selection methods form a \textbf{toolbox} where each tool is optimized to serve different goals.

When the goal is to make a persuasive chart feel \textbf{credible and complete}, we found that, surprisingly, nearly any guardrail was effective. \SemanticSim{} and \PercentileExemplars{} offered the largest gain, but were sometimes statistically indistinguishable from \RandomExemplars{}. Our results indicate that randomly-drawn items might be a ``false friend,'' offering credibility without providing any semantically-meaningful context.

On the other hand, when the goal is to help readers \textbf{accurately estimate the global rank} of a focal item, the best-performing methods are those that signal the data spread and typical trajectories. These methods include \Clusters{}, \Percentiles{}, and \PercentileExemplars{}. The \SemanticSim{} method, conversely, led to increased error.
These results confirm that participants leverage data-driven methods that surface distributional information but are biased by semantic context, which is entirely dependent on how the peer items perform relative to the rest of the data. An important insight is that \SemanticSim{} has the potential to mislead: while significantly improving credibility, it substantially biases global judgments, potentially exacerbating cherry-picking. 

\subsection{Context preferences are micro-level, not macro-level}

We found \textbf{no framing effect} on guardrail preference distributions in Task 2, as shown in Figure~\ref{fig:preferences}. Participants did not adjust their preferred chart based on the instructions or domain specifics. This may reflect the limitations of using prompts in an experimental setup to guide tasks, but also the fact that users form their preferences based on lower-level factors that are external to the task.

While we designed the guardrail sampling methods from a \textbf{macro-level} point of view (i.e., \textit{what data would make sense as context?} or \textit{how to best surface the distributional patterns in the dataset?}), the responses in Task 2 reveal that participants primarily base their selections on \textbf{micro-level} preferences. These include chart aesthetics, specific comparisons or insights they can make from the charts, or preferences for seeing a specific stock or country they expected to observe. Although many participants did cite broader reasons aligning their selections with the task they were performing, the quantitative results being consistent across scenarios indicate that this was primarily not the determining factor.

Overall, both the results of Task 2 suggest that participants prioritized \textbf{relevance and clarity} over the exact prompt. Taken together with the results of Task 1, we can conclude that when a chart surfaces the benchmarks people expect to see, provides insights people find interesting and non-contradictory, and is easy to read and aesthetically pleasing, people prefer it and trust it more. 

\section{System Design Implications}

Based on our studies, we distill the following design implications:
\begin{enumerate}
    \item \textbf{Prioritize Distributional Context for Accuracy:} When the objective is to help users accurately estimate the \textbf{global rank} of a focal item, only \textbf{data-driven guardrails} that encode distributional structure (\PercentileExemplars{}, \Percentiles{}, \Clusters{}) offer significant improvement.
    \item \textbf{Use Semantic Context for Trust and Relevance:} When the goal is to maximize \textbf{user trust and perceived context completeness}, any guardrail is effective. Use \SemanticSim{} to satisfy user preference for \textbf{relevance}, but be aware: it can reduce accuracy for global judgments by misleadingly framing the data within a non-representative subgroup.
    \item \textbf{Explain the Contextual Choice:} Since user preferences are often pre-determined and independent of the task or domain, the system should \textbf{explicitly surface and explain} why the specific guardrail items were chosen. This is necessary to bridge the gap between user-preferred micro-level cues and the macro-level goal of the guardrail (task relevance, distributional context).
    \item \textbf{Adopt a Goal-Oriented Guardrail Toolbox and Mitigate Risk:} Implement a suite of context selection methods and deploy the method that best serves the system's use cases. Crucially, recognize that seemingly successful and easiest to implement methods like \RandomExemplars{} can be a ``false friend,'' offering credibility without meaningful context and potentially masking important data patterns.
\end{enumerate}

\section{Limitations and Future Work}

Our study focuses on two real-world domains (COVID-19 cases and stocks) and examines a small set of non-extreme items. Future work could extend guardrail evaluations to other issues such as climate or election data, vary focal item rank to contrast high- and low-performing groups, and test whether effects depend on how consequential or familiar a domain is to participants. 

We also fixed the number of contextual lines to five, following prior evidence that this strikes a balance between trust and performance~\cite{padilla_multiple_2022}. Subsequent research can map how trust, context, and accuracy shift as the number and saliency of contextual lines change, identifying task-specific sweet spots. Our study also kept the number of focal items fixed, and future work could explore methods to accommodate multiple selections. Most methods are independent of the focal item and should thus scale easily to more focal points, whereas \SemanticSim{} may require selecting guardrails that are similar to each focal item. As the number of focal items grows, it may also become advantageous to reduce the number of guardrails to address visual clutter.

Our design used concise, persuasive captions, but alternative framings---such as more adversarial narratives or neutral descriptions---may shift perceptions and outcomes. Future work may explore this narrative space, including how guardrails interact with varying framing strengths. Finally, because our US and UK crowd-sourced sample reflects a specific cultural lens, generalizability may differ across contexts, as notions of ``appropriate'' or useful comparators can vary by culture and experience. 

\section{Conclusion}

Guardrails embed a set of contextual comparators into data explorers, better preparing basic charts for the various tasks people use them for. We proposed five guardrail sampling strategies and confirmed that all significantly improve trust and context, while data-driven methods improve accuracy in rank judgments. Through a crowd-sourced evaluation across multiple framings, tasks, and scenarios, our paper presents actionable recommendations for data explorer platform governance. Specifically, platforms should surface semantically-comparable peers for maximizing credibility, and select percentile-based exemplars for achieving precise judgment. In summary, our strategies allow data explorer platforms to move beyond simply displaying data to actively contextualizing decisions and fostering reliable interpretation for the general public.

\section{Acknowledgments}

This work is supported by the National Science Foundation (CNS 2213756).

\bibliographystyle{eg-alpha-doi} 
\bibliography{references}                   

\clearpage
\appendix
\setcounter{page}{1}
\input{appendix}

\end{document}

%% file: appendix.tex
\renewcommand{\thefigure}{A\arabic{figure}}
\setcounter{figure}{0}

\renewcommand{\thetable}{A\arabic{table}}
\setcounter{table}{0}

\section{Stimuli, Datasets, and Materials}
\label{app:stimuli}
To facilitate review and replication, we provide direct links to the deployed study:
\begin{itemize}
  \item COVID-19 scenario: \url{https://vdl.sci.utah.edu/guardrail-samples-study/stage-1-covid}
  \item Stocks scenario: \url{https://vdl.sci.utah.edu/guardrail-samples-study/stage-1}
  \item Condition explorer (all guardrail variants used in-study, plus prototypes): \url{https://vdl.sci.utah.edu/guardrail-samples-study/sandbox}
\end{itemize}

\section{LLM Prompts and Final Selection Sets for Semantic Similarity (LLM) Condition}
\label{app:prompts}

\subsection{COVID-19 prompt and selections}
\noindent\textbf{Prompt.}
\begin{quote}\small
You are curating contextual comparisons for a public data-exploration platform that visualizes COVID-19 cumulative cases per million. The goal is to help people make better sense of charts, surface missing context, and ultimately support better decisions and a more holistic understanding of the metric. For each highlighted country, select five other countries as meaningful comparisons to co-plot as auxiliary lines. Consider geographic proximity (e.g., immediate neighbors), similar stages of economic development, comparable demographics and urbanization, and health-system capacity. Avoid random picks, duplicates, and microstates unless the anchor is one; relax constraints only as needed, and break ties by geographic proximity, then alphabetically.
\end{quote}

\paragraph{Final selection sets used in-study.}
\begin{itemize}
  \item \textbf{Greece}: Italy, Spain, Portugal, Cyprus, Croatia
  \item \textbf{Germany}: France, Netherlands, Austria, Sweden, Denmark
  \item \textbf{Belarus}: Russia, Ukraine, Kazakhstan, Moldova, Serbia
  \item \textbf{Norway}: Sweden, Denmark, Finland, Iceland, Netherlands
\end{itemize}

\subsection{Stocks prompt and selections}
\noindent\textbf{Prompt.}
\begin{quote}\small
You are curating contextual comparisons for a public data-exploration platform that visualizes S\&P 500 stock price performance (percentage change). The goal is to help people make better sense of charts, surface missing context, and ultimately support better decisions and a more holistic understanding of the metric. For each highlighted stock, select five other stocks as meaningful comparisons to co-plot as auxiliary lines. Consider industry proximity (e.g., same or adjacent GICS industry or sub-industry), similar market capitalization, comparable growth/volatility profiles, and operating capacity. Avoid random picks, duplicates, and non-S\&P 500 stocks; relax constraints only as needed, and break ties by industry proximity, then alphabetically by ticker.
\end{quote}

\paragraph{Final selection sets used in-study.}
\begin{itemize}
  \item \textbf{COR}: MCK, CAH, CVS, WBA, HSIC
  \item \textbf{CHD}: PG, CL, KMB, CLX, EL
  \item \textbf{TEL}: APH, KEYS, GLW, HUBB, ETN
  \item \textbf{VZ}: T, TMUS, CMCSA, CHTR, AMT
\end{itemize}


\section{Participant Demographics}

We recruited a total of $N = 487$ participants after preregistered exclusions.
Participants had a mean age of 41.74 years (SD = 12.44; median = 39.50).

\begin{figure}[t]
    \centering
    \includegraphics[width=\linewidth]{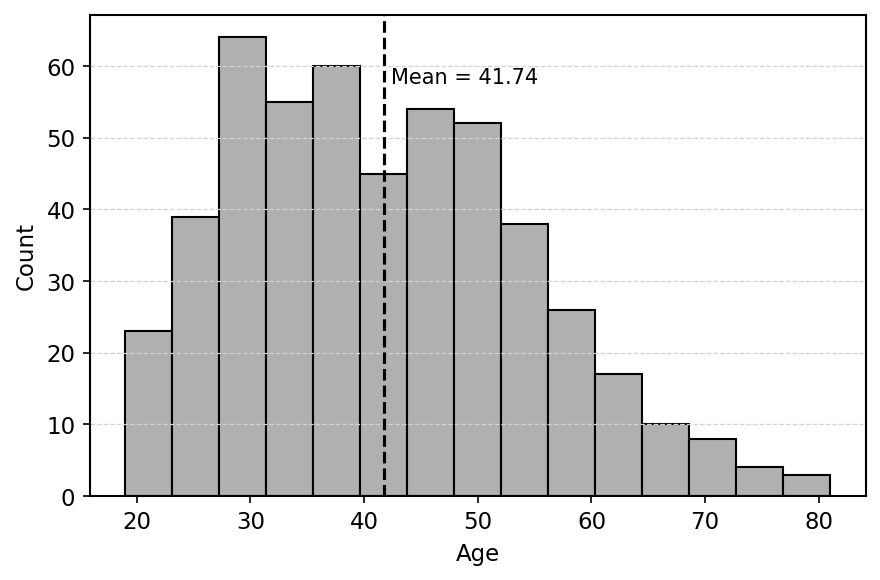}
    \caption{Age distribution of participants.}
    \label{fig:age_dist}
\end{figure}

\begin{figure}[t]
    \centering
    \includegraphics[width=\linewidth]{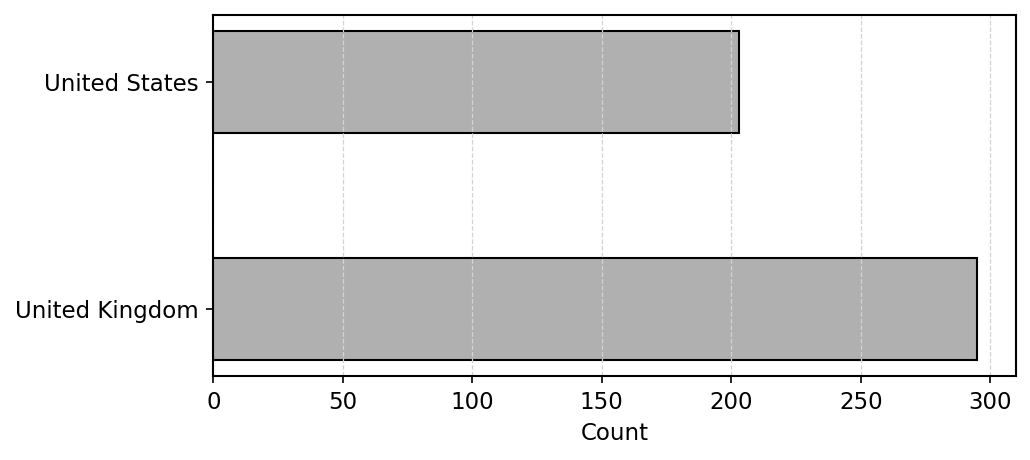}
    \caption{Distribution of participants by country of residence.}
    \label{fig:country_dist}
\end{figure}

\begin{figure}[t]
    \centering
    \includegraphics[width=\linewidth]{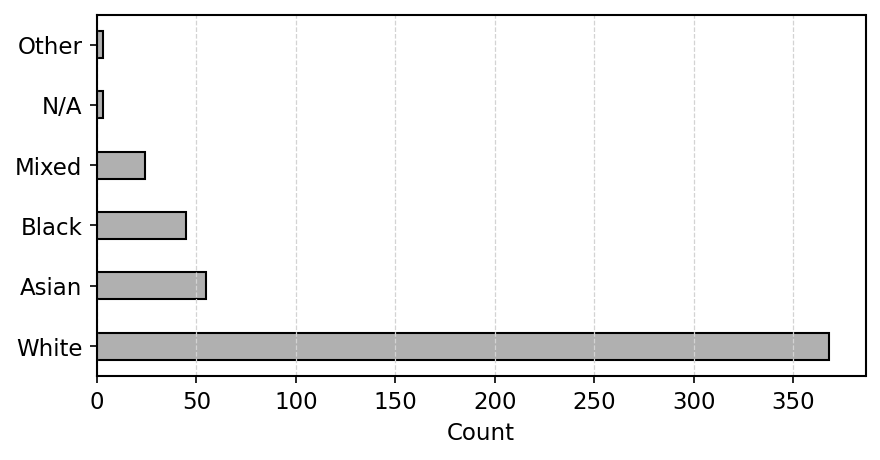}
    \caption{Distribution of participants by self-reported ethnicity.}
    \label{fig:ethnicity_dist}
\end{figure}

\begin{figure}[t]
    \centering
    \includegraphics[width=\linewidth]{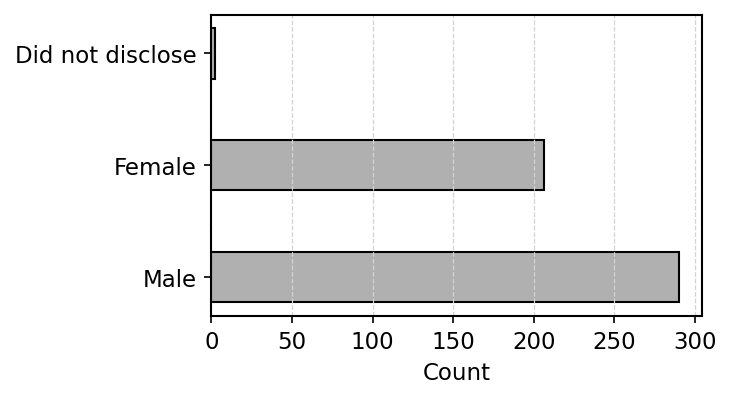}
    \caption{Distribution of participants by self-reported sex.}
    \label{fig:sex_dist}
\end{figure}

Figures~\ref{fig:age_dist}--\ref{fig:sex_dist} provide detailed visual
summaries of the demographic distributions.
\section{Analysis Scripts}
\label{app:analysis}
Full analysis notebooks (model fits, bootstrap CIs, and figure generation) are available here: \\
\url{https://colab.research.google.com/drive/1YyBcm-NKntIQLPzOI-K49cCf6YAfwKqg?usp=sharing}

\section{Artifacts}
\label{app:repro}
\begin{itemize}
  \item Study code repository: \url{https://github.com/visdesignlab/guardrail-samples-study}
  \item COVID-19 dataset: \url{https://github.com/visdesignlab/guardrail-samples-study/blob/data-v1.0/public/stage-1-covid/data/clean_data.csv}
  \item Stocks dataset: \url{https://github.com/visdesignlab/guardrail-samples-study/blob/data-v1.0/public/stage-1/data/sp500_stocks.csv}
\end{itemize}


\section{Additional Tables, Figures and Screenshots}
\clearpage

\clearpage

\begin{table*}[t]
\centering
\caption{Exact $p$-values for preregistered hypothesis tests.}
\label{tab:exact_pvalues}
\begin{tabular}{lllll}
\toprule
Hypothesis & Measure & Scenario & $\chi^2$ (df) & Exact $p$ \\
\midrule

\multicolumn{5}{l}{\textbf{H1: Guardrails vs. No Guardrail}} \\
\midrule
H1 & Trust    & COVID  & 16.457 (1)  & $4.9771 \times 10^{-05}$ \\
H1 & Trust    & Stocks & 22.327 (1)  & $2.2998e \times 10^{-06}$ \\

H1 & Accuracy & COVID  & 15.704 (1)  & $7.4049 \times 10^{-05}$ \\
H1 & Accuracy & Stocks & 12.026 (1)  & $5.2451 \times 10^{-04}$ \\

H1 & Context  & COVID  & 100.171 (1) & $1.3978 \times 10^{-23}$ \\
H1 & Context  & Stocks & 113.079 (1) & $2.0734 \times 10^{-26}$ \\

\midrule

\multicolumn{5}{l}{\textbf{H2: Context-rich vs. Random}} \\
\midrule
H2 & Trust    & COVID  & 1.8130 (1)  & $1.7813 \times 10^{-01}$ \\
H2 & Trust    & Stocks & 0.3240 (1)  & $5.6916 \times 10^{-01}$ \\

H2 & Accuracy & COVID  & 3.1100 (1)  & $7.7793 \times 10^{-02}$ \\
H2 & Accuracy & Stocks & 11.176 (1)  & $8.2882 \times 10^{-04}$ \\

H2 & Context  & COVID  & 1.7780 (1)  & $1.8237 \times 10^{-01}$ \\
H2 & Context  & Stocks & 0.3470 (1)  & $5.5556 \times 10^{-01}$ \\

\midrule

\multicolumn{5}{l}{\textbf{H3: Task Framing}} \\
\midrule
H3 & Framing  & COVID  & 0.5050 (3)  & $9.1782 \times 10^{-01}$ \\
H3 & Framing  & Stocks & 0.5500 (3)  & $9.0779 \times 10^{-01}$ \\

\bottomrule
\end{tabular}
\end{table*}

\clearpage

\begin{figure*}[]
  \centering
  \includegraphics[width=\linewidth]{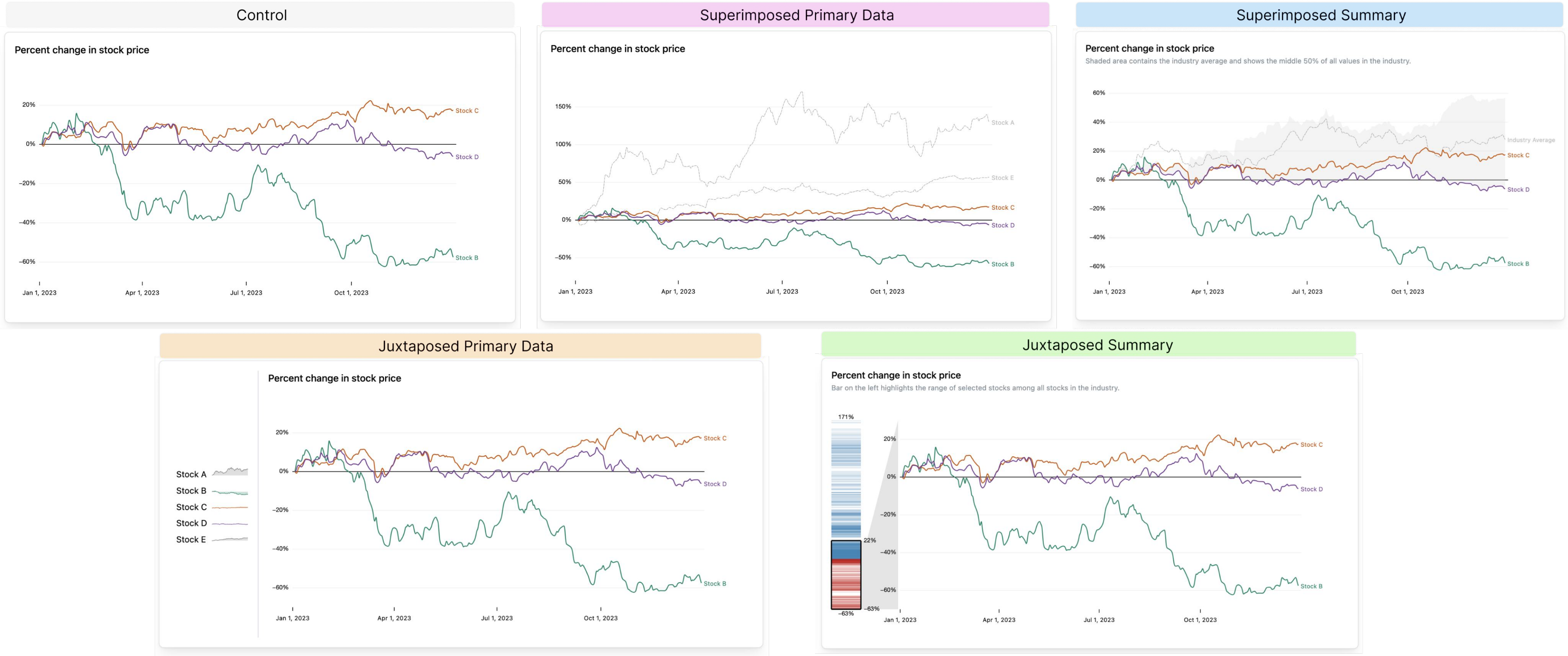}
  \caption{Example of the guardrail representation types identified by previous work~\cite{lisnic_visualization_2025}. We build upon these results, identifying sampling strategies for the most effective Superimposed Primary Data condition.}
  \label{fig:old-guardrails}
\end{figure*}

\clearpage

\begin{figure*}[]
  \centering
  \includegraphics[width=.92\textwidth]{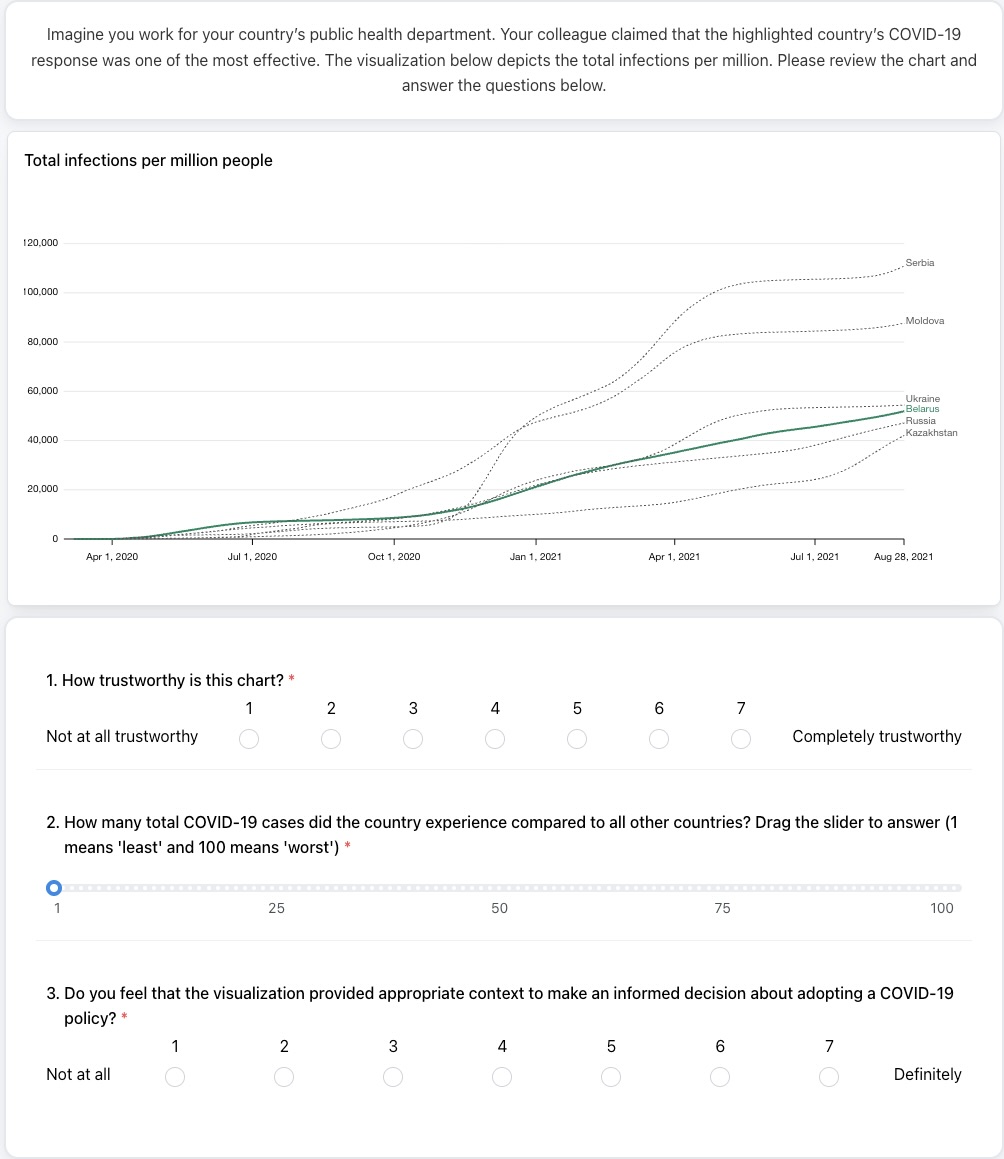}
  \caption{Screenshot of the Task 1 stimulus and survey in the COVID-19 scenario.}
  \label{fig:app-covid-1}
\end{figure*}

\begin{figure*}[]
  \centering
  \includegraphics[width=.92\textwidth]{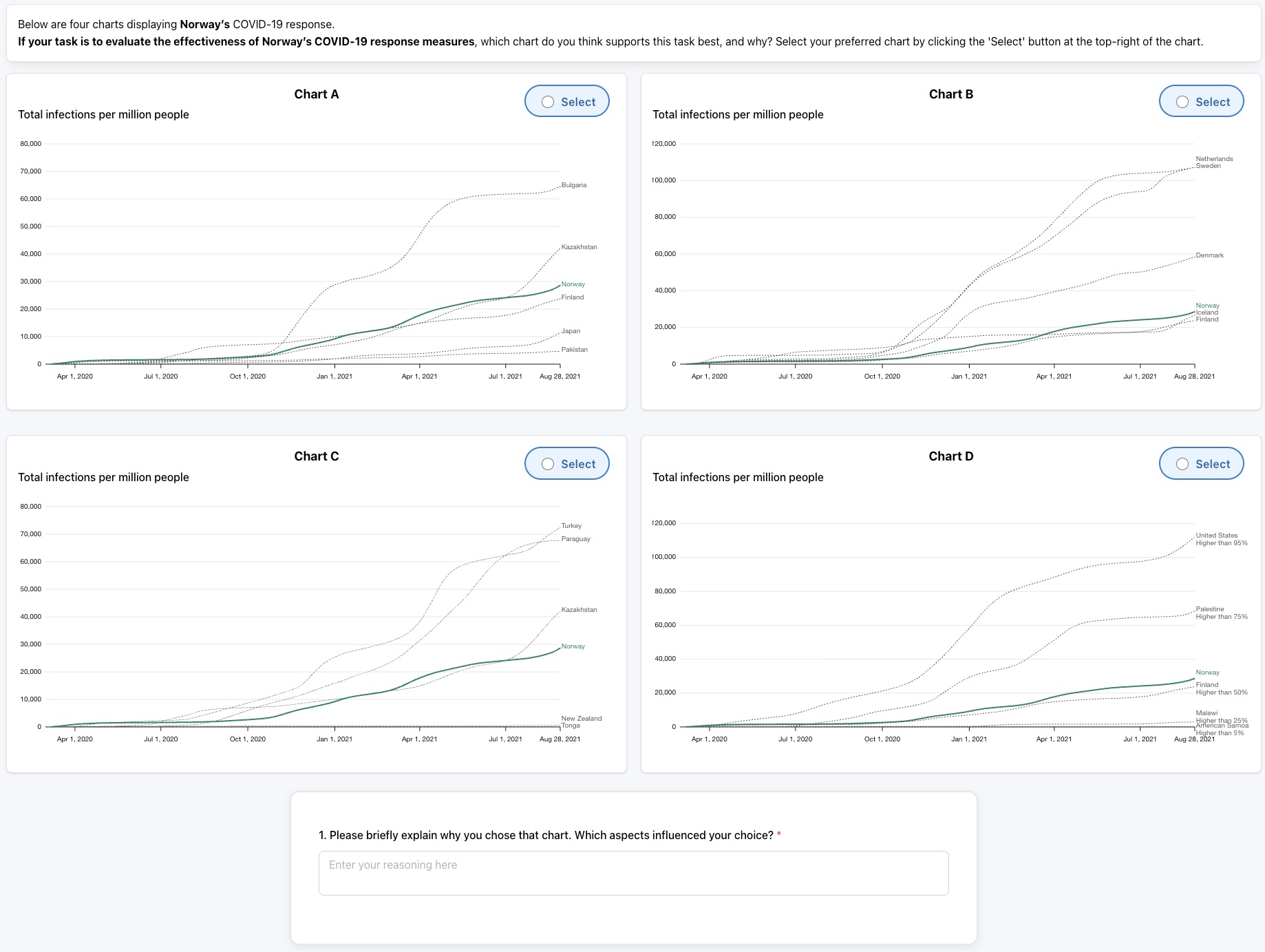}
  \caption{Screenshot of the Task 2 stimulus and survey in the COVID-19 scenario.}
  \label{fig:app-covid-2}
\end{figure*}

\begin{figure*}[]
  \centering
  \includegraphics[width=.92\textwidth]{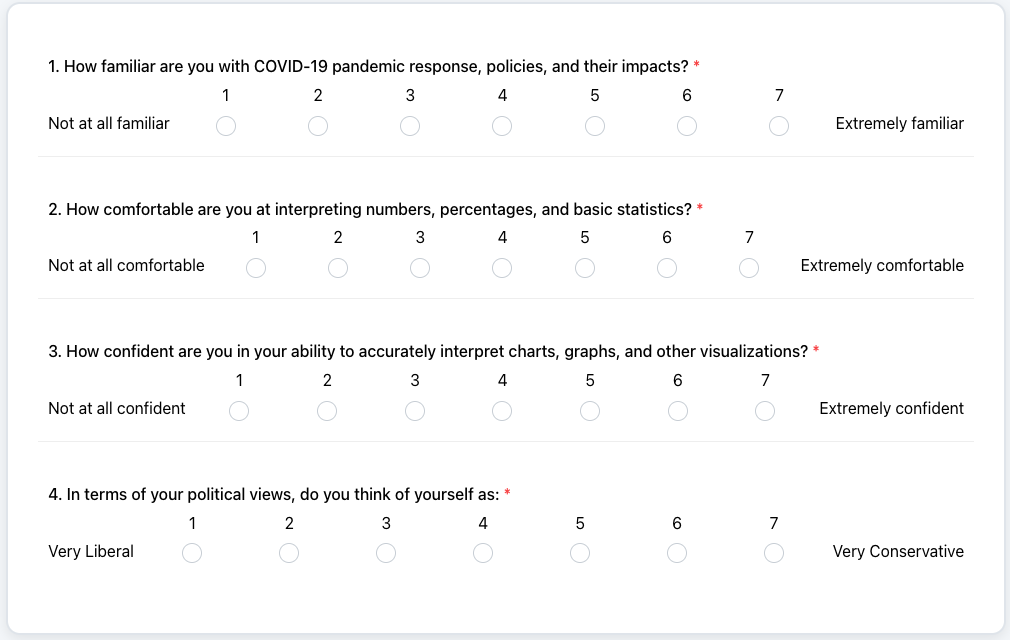}
  \caption{Screenshot of the post-study survey in the COVID-19 scenario.}
  \label{fig:app-covid-survey}
\end{figure*}

\begin{figure*}[]
  \centering
  \includegraphics[width=.92\textwidth]{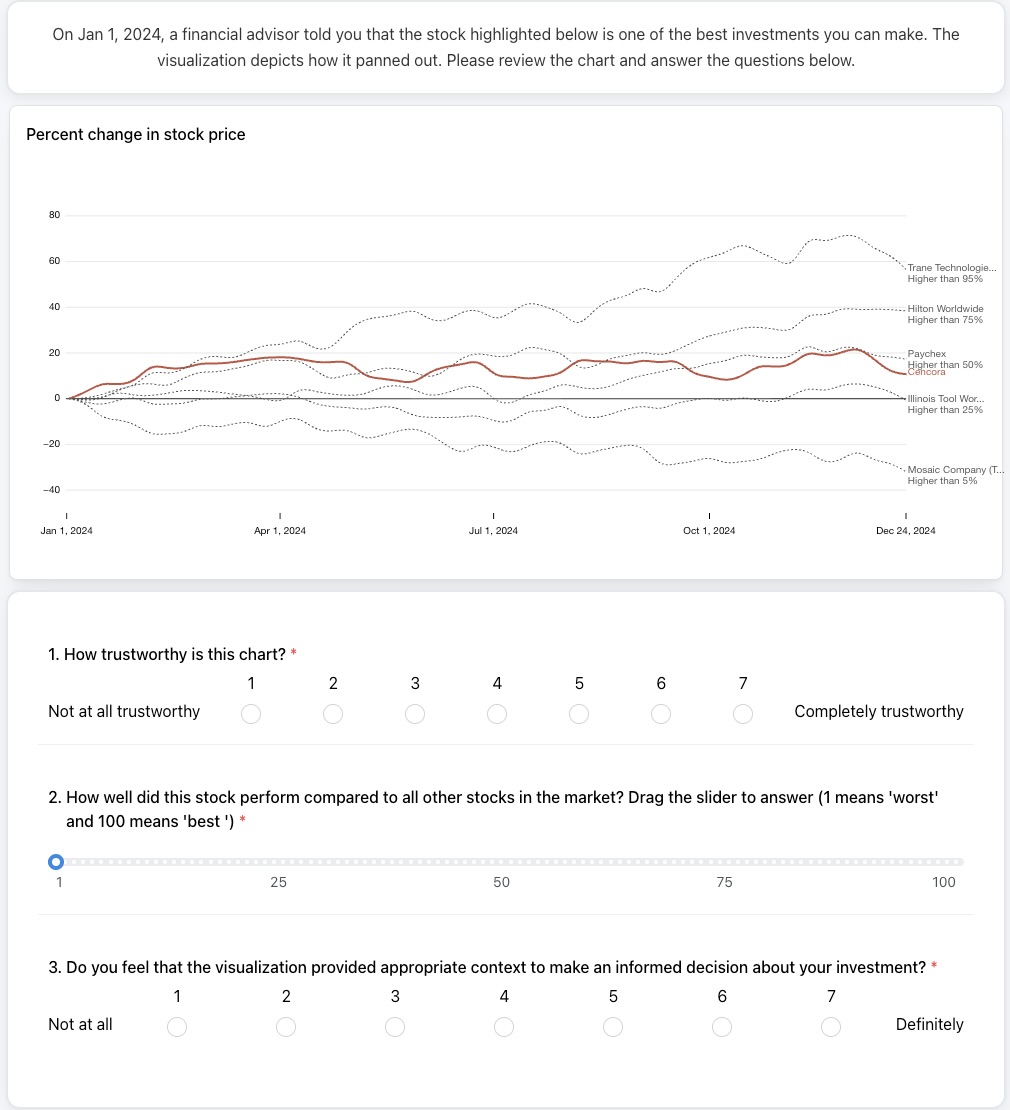}
  \caption{Screenshot of the Task 1 stimulus and survey in the Stocks scenario.}
  \label{fig:app-stocks-1}
\end{figure*}

\begin{figure*}[]
  \centering
  \includegraphics[width=.92\textwidth]{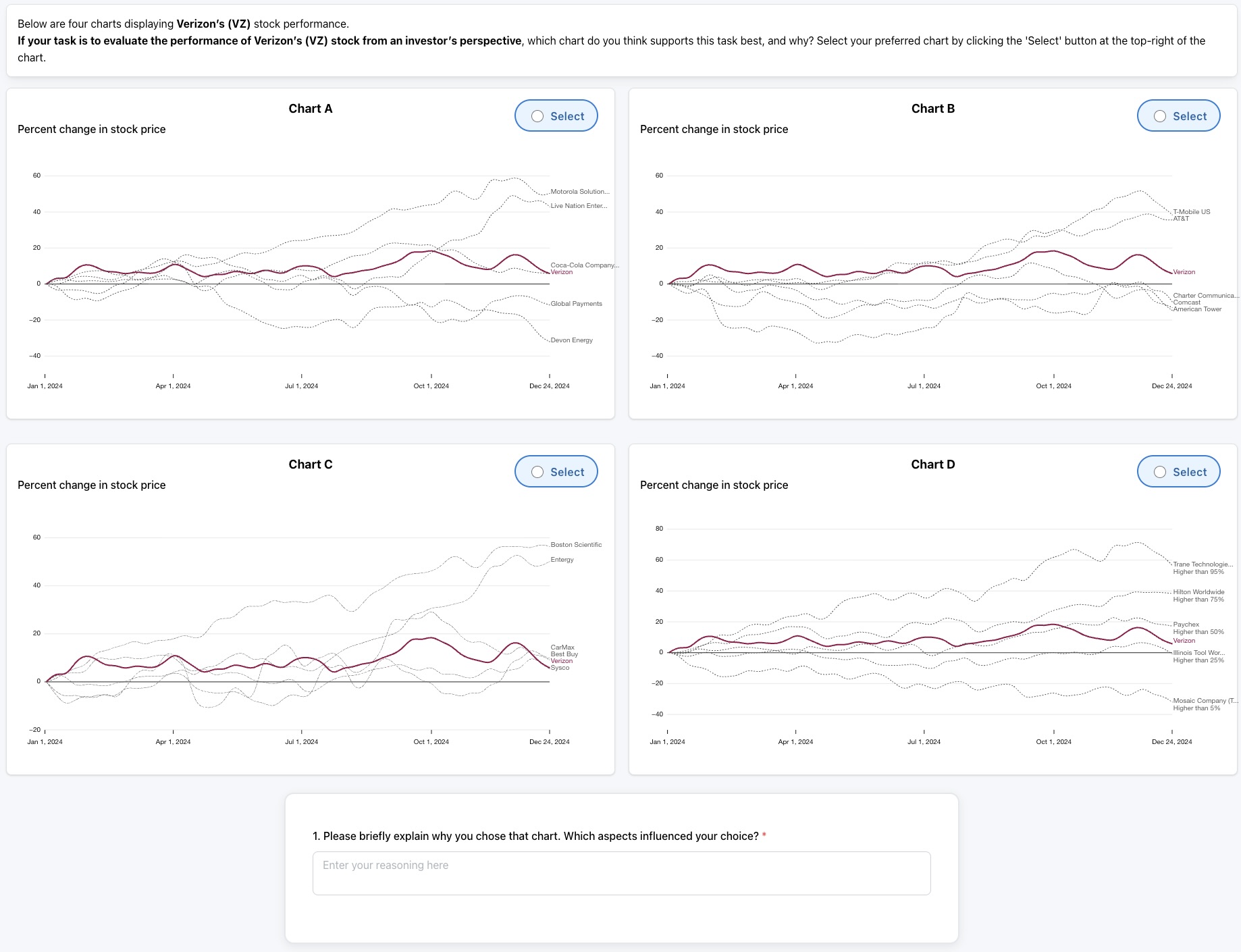}
  \caption{Screenshot of the Task 2 stimulus and survey in the Stocks scenario.}
  \label{fig:app-stocks-2}
\end{figure*}

\begin{figure*}[]
  \centering
  \includegraphics[width=.92\textwidth]{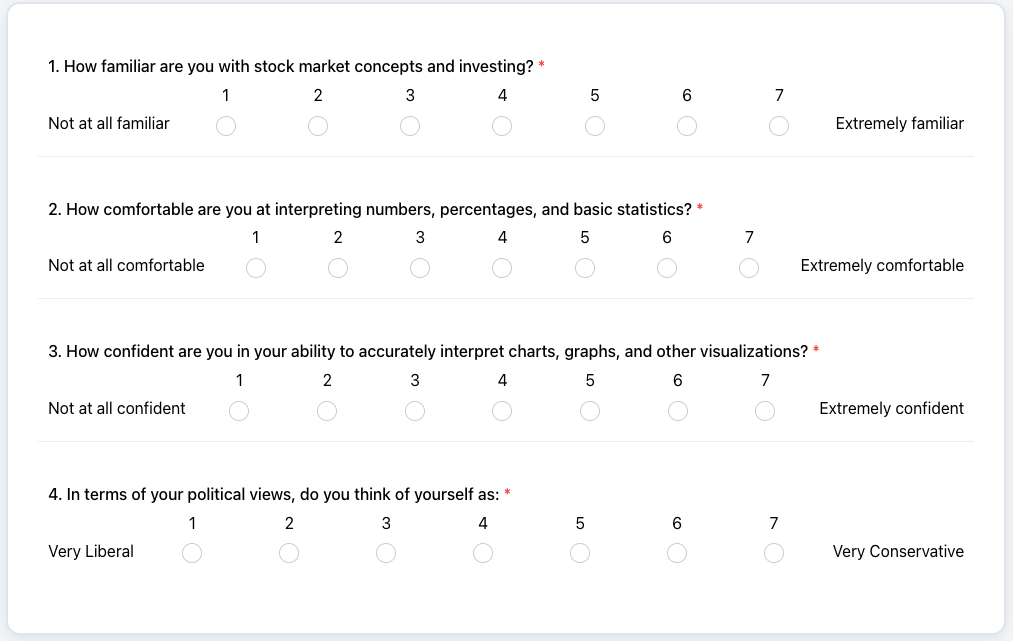}
  \caption{Screenshot of the post-study survey in the Stocks scenario.}
  \label{fig:app-stocks-survey}
\end{figure*}

\clearpage
\begin{figure*}[]
  \centering
  \includegraphics[width=\linewidth]{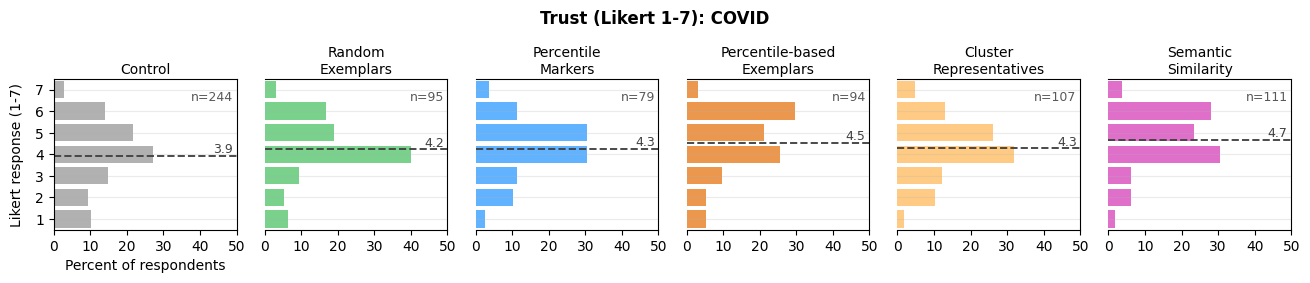}
  \caption{Distributions of Trust responses by guardrail selection method in the COVID-19 scenario in Task 1.}
  \label{fig:covid-trust}
\end{figure*}
\begin{figure*}[]
  \centering
  \includegraphics[width=\linewidth]{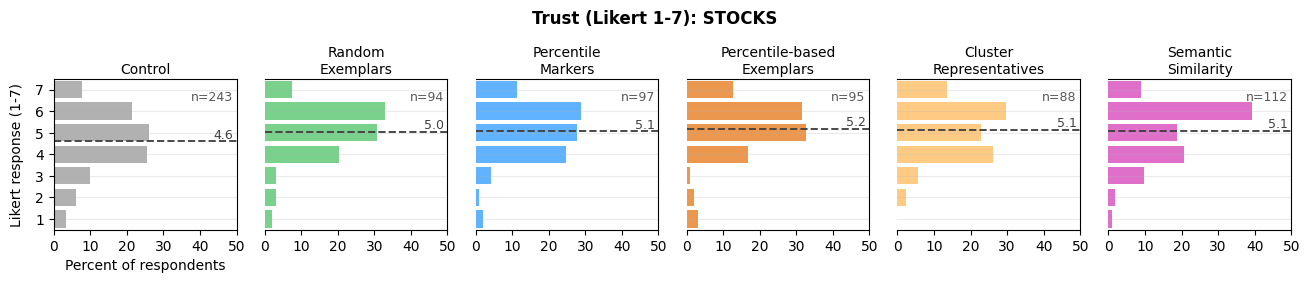}
  \caption{Distributions of Trust responses by guardrail selection method in the Stocks scenario in Task 1.}
  \label{fig:stock-trust}
\end{figure*}

\clearpage
\begin{figure*}[]
  \centering
  \includegraphics[width=\linewidth]{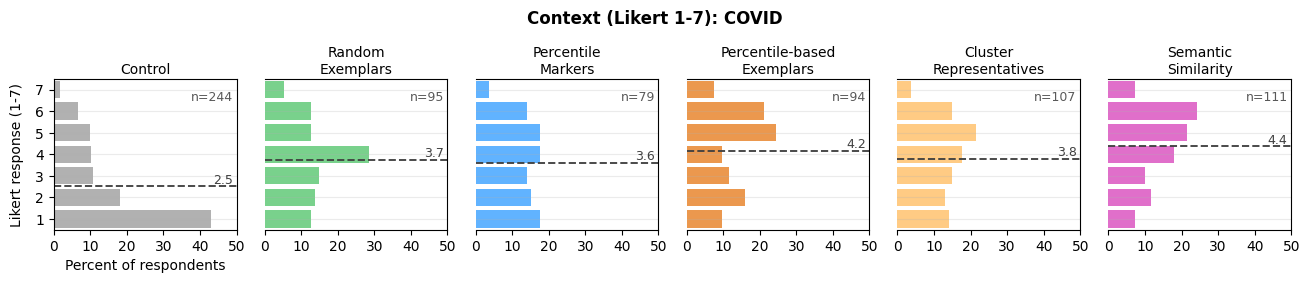}
  \caption{Distributions of Context responses by guardrail selection method in the COVID scenario in Task 1.}
  \label{fig:covid-context}
\end{figure*}

\begin{figure*}[]
  \centering
  \includegraphics[width=\linewidth]{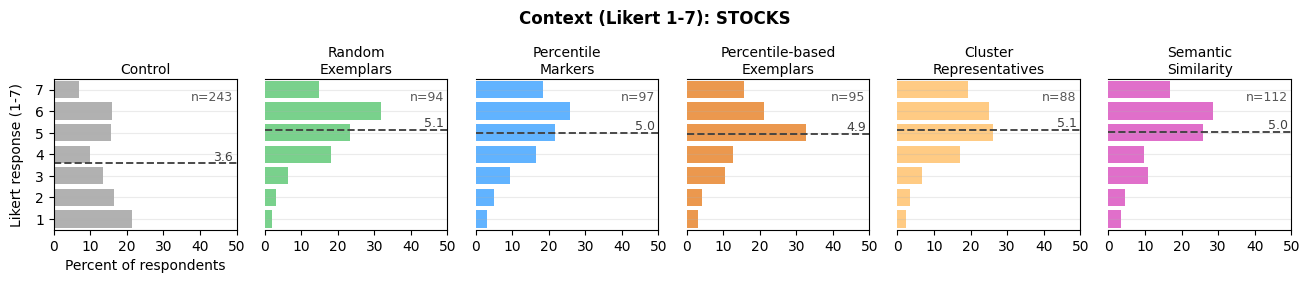}
  \caption{Distributions of Trust responses by guardrail selection method in the Stocks scenario in Task 1.}
  \label{fig:stock-context}
\end{figure*}